\documentclass{aa}  

\usepackage{graphicx}
\usepackage{txfonts}
\usepackage{hyperref}
\usepackage{xcolor}

\usepackage{subfigure}

\usepackage{amsmath}
\usepackage{algorithm}
\usepackage[noend]{algpseudocode}
\usepackage{mathrsfs} 
\usepackage{dsfont}

\newcommand{\norm}[1]{\left\lVert#1\right\rVert}

\graphicspath{{./}{figures/}}

\begin{document}

\title{Multi-scale and Multi-directional VLBI Imaging with CLEAN}

\author{H. Müller
        \inst{1}
        \and
        A.P. Lobanov\inst{1}
        }

\institute{Max-Planck-Institut für Radioastronomie,
        Auf dem Hügel 69, Bonn, 53121, Germany\\
        \email{hmueller@mpifr-bonn.mpg.de}, \email{alobanov@mpifr-bonn.mpg.de}
        }
        
        
\date{Received September 15, 1996; accepted March 16, 1997}

\abstract
{Very long baseline interferometry (VLBI) is a radio-astronomical technique in which the correlated signal from various baselines is combined into an image of highest angular resolution. Due to sparsity of the measurements, this imaging procedure constitutes an ill-posed inverse problem. For decades the CLEAN algorithm was the standard choice in VLBI studies, although having some serious disadvantages and pathologies that are challenged by the requirements of modern frontline VLBI applications.}
{We develop a novel multi-scale CLEAN deconvolution method (DoB-CLEAN) based on continuous wavelet transforms that address several pathologies in CLEAN imaging. We benchmark this novel algorithm against CLEAN reconstructions on synthetic data and reanalyze BL Lac observations of RadioAstron with DoB-CLEAN.}
{DoB-CLEAN approaches the image by multi-scalar and multi-directional wavelet dictionaries. Two different dictionaries are used. Firstly, a difference of elliptical spherical Bessel functions dictionary fitted to the uv-coverage of the observation that is used to sparsely represent the features in the dirty image. Secondly, a difference of elliptical Gaussian wavelet dictionary that is well suited to represent relevant image features cleanly. The deconvolution is performed by switching between the dictionaries.}
{DoB-CLEAN achieves super-resolution compared to CLEAN and remedies the spurious regularization properties of CLEAN. In contrast to CLEAN, the representation by basis functions has a physical meaning. Hence, the computed deconvolved image still fits the observed visibilities, opposed to CLEAN.}
{State-of-the-art multi-scalar imaging approaches seem to outperform single-scalar standard approaches in VLBI and are well suited to maximize the extraction of information in ongoing frontline VLBI applications.} 

\keywords{Techniques: interferometric - Techniques: image processing - Techniques: high angular resolution - Methods: numerical - Galaxies: jets - Galaxies: nuclei}

\maketitle

\section{Introduction}\label{sec: intro}
Very long baseline interferometry (VLBI) is a radio-interferometric technique that achieves unmatched angular resolution. An array of single-dish antennas form together an instrument with angular resolution determined by the wavelength and longest separation between two antennas in the array \citep{Thompson1994}. The signal recorded at each antenna pair is correlated. The correlation product (visibility) is proportional to the Fourier-transform of the true sky-brightness distribution (van Cittert-Zernike theorem) where the spatial frequency is specified by the baseline separating the two antennas recording. In principle the true image could be revealed from a complete sampling of the uv-space by an inverse Fourier transform. However, since an interferometer is a sparse array of single antennas with a limited number of baselines, the coverage of Fourier coefficients (uv-coverage) is often sparse and has significant gaps \citep{Thompson1994}. This makes imaging, i.e. the procedure of creating an image from the correlated antenna outputs, an ill-posed inverse problem.

The imaging problem (inverse Fourier transform from sparsely sampled data) is often expressed equivalently as a deconvolution problem, i.e. the dirty image (inverse Fourier transform of visibilities) is modeled as the convolution of the dirty beam (inverse Fourier transform of mask) and the true sky brightness distribution. \citep{Thompson1994}

CLEAN and its variants \citep{Hogbom1974, Clark1980, Schwab1984} have been the standard in VLBI imaging for decades and still remain widely used. CLEAN models the image iteratively as a set of point sources: CLEAN searches for the position of the maximum in the residual image, stores the intensity and the position in a list of delta-components, and updates the residual by subtracting the rescaled and shifted dirty beam from the residual image. Despite the general success of CLEAN in VLBI applications, there is a number of known issues by now: CLEAN is less precise than recently developed regularized maximum likelihood (RML) methods \citep{Akiyama2017, Akiyama2017b, Chael2018, eht2019d, Mueller2022} and Bayesian approaches \citep{Arras2020}, in particular if the true sky brightness distribution is uniform and extended, it provides poorer resolution, and relies on manual input from the astronomer performing the imaging to achieve convergence to the true solution. Moreover, the sequential nature inherent to CLEAN makes CLEAN slow compared to modern optimization algorithms that were developed in an environment of parallel CPU computing facilities.

From a theoretical point of view CLEAN is inadequate. An imaging procedure needs to satisfy two basic requirements. Firstly, the final image needs to fit the observed visibilities. Secondly, among all possible solutions that fit the data (i.e. among the kernel spanned by the convolution with the dirty beam) the imaging procedure should select the image that is most reasonable, i.e. that interpolates the gaps in the uv-coverage in the most reasonable way. CLEAN can only achieve one of these goals simultaneously. CLEAN separates between a model (the list of delta-components) that fits the observed data and the final image (the model convolved with a clean beam) that is thought to be a reasonable approximation to the true sky brightness distribution. However, strictly speaking, the final image that CLEAN produces in VLBI (and that is used in further studies) does not provide a reasonable data fit anymore.

In fact, the regularizing property of CLEAN is questionable as well. While CLEAN typically provides decent fits for the uv-tracks that were observed, the (typically not plotted) fit in the gaps in the uv-coverage is sometimes clearly unphysical, we will discuss this attribute in more detail in Sec. \ref{sec: synthetic_test}. A more thorough imaging approach is needed that takes the distribution of gaps in the uv-coverage in account and provides more control over the non-measured Fourier coefficients.

Most of these issues are caused by CLEAN modeling the image as a sequence of delta components which is inadequate to describe extended image features in real astronomical images. A possible solution is the use of multi-scalar algorithms that model the image as a set of extended basis functions of different scales \citep{Wakker1988, Starck1994, Bhatnagar2004, Cornwell2008, Rau2011, Mueller2022}. While this is a great step forward in imaging, MS-CLEAN methods have not been widely adopted in frontline VLBI applications in the past. This is because the selection of suitable basis functions greatly affects the fitting procedure as various scales are sensitive to various parts of the uv-coverage and do not necessarily solve the problem of missing regularization in CLEAN, i.e. the unphysical fits in the gaps of the uv-coverage. To also address this problem of missing regularization, we propose a more data-driven approach here: the basis functions are selected in a way that they fit to the uv-coverage, i.e. that they define masks in the Fourier domain that separate between visibilities covered by observations and visibilities that are not covered by observations (gaps in the uv-coverage). The features from the latter should be suppressed during imaging, i.e. the unphysical fit in the gaps occurring during CLEAN should be smoothed/regularized. As the uv-coverage of an observation is typically not circularly symmetric, we propose (for the first time) not only a multi-scalar, but also a multi-directional set of basis functions (dictionary).

In this way our procedure allows for a more thorough separation between reliable image information, i.e. image features introduced by regions in the Fourier domain that are covered by data, and `invisible distributions', i.e. image features that are most sensitive to regions of the uv-coverage that are not covered by observations. This is well needed to match our second basic requirement for an imaging algorithm for frontline VLBI arrays, i.e. that among all possible solutions the one that is most physical (regularized) should be selected.

We present in this paper how to construct a suitable multi-scalar and multi-directional dictionary for imaging and how this dictionary can be implemented in a CLEAN like algorithm, called DoB-CLEAN (difference of elliptical Bessel functions CLEAN), that fits in the normal workflow that radio astronomers are used to. 


\section{Theory} \label{sec: theory}

\subsection{Background}
A radio interferometer observes a source with all antennas available in the array at the same time. The source typically appears point-like per antenna in the constructed array. The interferometric observation however reveals image features at much greater resolution. We denote the (incoherent) sky-brightness distribution of the source by $I(l, m)$. Here $l$ and $m$ denote spatial on-sky coordinates. The recorded signals are correlated for each antenna pair at a fixed time. The antenna pair is specified by a corresponding separation vector $(u,v)$ (spatial frequencies in units of wavelengths), which is called baseline. While the Earth rotates during the time of observation, the projected baselines vary as well, leading to the typical elliptical tracks in the uv-coverage. Described by the van-Cittert-Zernike theorem (assuming the small-field approximation and a flat wavefront), the correlation product at a single baseline is the Fourier coefficient of the true sky brightness distribution at this baseline \citep{Thompson1994}:
\begin{align}
    \mathcal{V} (u, v) = \int \int I(l, m) e^{-2 \pi i (l u + m v)} dl dm \,,  \label{eq: vis}
\end{align}
These Fourier coefficients are called visibilities. 

Imaging is the problem of recovering the on-sky distribution $I$ from the measured complex visibilities $\mathcal{V}$. From a full sample of the uv-domain, this could be achieved by an (gridded) inverse Fourier transform. However, every antenna pair at a a particular instance in time gives rise to only one Fourier coefficient. Hence, the limited number of available antennas and the limited amount of observation time allows for only a very sparse coverage of the uv-domain. 

For imaging with CLEAN \citep{Hogbom1974}, Eq. \eqref{eq: vis} is equivalently reformulated as a deconvolution problem. The observed visibilities are gridded on a regular grid and possibly weighted (e.g. by baseline-dependent signal-to-noise ratio and, in the case of uniform weighting, by the number of data points per cell). The gridding cells corresponding to unmeasured Fourier coefficients are set to zero. The dirty image $I^\mathrm{D}$ is now defined as the inverse Fourier transform of the gridded (and weighted) observed visibilities. Furthermore, the dirty beam $B^D$ is the response to a synthetic point-source, i.e. the inverse Fourier transform of the gridding (and weighting) alone. It is:
\begin{align}
    I^\mathrm{D} = B^D * I. \label{eq: dirty_map}
\end{align}
The imaging problem is now translated in a deconvolution problem. The dirty image and the dirty beam contain significant sidelobes that are caused by the gaps in the uv-coverage, i.e. the cells in Fourier domain that are initialized with zero during gridding. These sidelobes are `cleaned', i.e. suppressed,  by deconvolution. Hence, the deconvolution process can also be understood as an approach to interpolate the observed measured visibilities to the gaps.

Among the sparsity of the observed Fourier coefficients, the imaging procedure has to deal with further complications: scale-dependent thermal noise on different baselines and direction-independent calibration issues. The former complication is addressed by weighting the visibilities by their thermal noise level. The latter complication is factored in station-based multiplicative gains. In particular, the relative phase is often unknown in VLBI imaging. Station based gains are corrected by gain-self-calibration loops alternating with deconvolution iterations. In principle, also more complex calibration errors could occur that cannot be factored in station-based gains at all.

\subsection{CLEAN} \label{ssec: clean}
CLEAN directly solves the deconvolution in Eq. \eqref{eq: dirty_map} by iteratively subtracting the dirty beam from the residual. Classical CLEAN \citep{Hogbom1974} approaches the image as a sequence of point sources. Hence, once the position of a new component is found, the dirty beam is shifted to this position and rescaled to the intensity in the residual image at that location multiplied with some gain parameter. The residual image is updated by subtracting the shifted and rescaled dirty beam. The list of delta-components constitutes the model that CLEAN computes to fit the observed visibilities.

It is crucial for CLEAN to find  a proper location of the next component. This is handled mostly manually by the astronomer by specifying search-windows for the next components. This procedure has proven successful, in particular in the presence of calibration errors. However, the iterative windowing, flagging and self-calibration lacks reproducibility. Within the specified window, the location of the next component is found by the location of the peak in the current residual image. However, this is only approximately correct. If the assumption behind CLEAN, i.e. that the true sky brightness distribution is modeled by a sum of point-sources, were true and we would ignore thermal noise for one moment, the current residual ($I^D$) could be envisioned as the convolution of the dirty beam $B^D$ with the sum of point sources that are unmodeled by CLEAN until this step ($\{ \delta_l \}$ with intensities $a_l$):
\begin{align}
    I^{D} = \sum_l a_l B^D * \delta_l. \label{eq: point-sources}
\end{align}
The most efficient selection criterion would be to find the largest of these unmodeled point-sources, i.e. the largest $a_l$. CLEAN takes the largest peak in the residual instead. This might not always be the optimal choice since overlapping sidelobes from different emission features can suppress real emission, and can create a false source when the sidelobes constructively add. In practice, this subtle difference however was not found to cause problems. However, we like to note that the new multi-scale CLEAN (DoB-CLEAN) algorithm that we propose in Sec. \ref{sec: algorithm} will be based on the same assumption, see Sec. \ref{ssec: selection}.

After the final CLEAN-iteration, the list of delta components is typically convolved with a clean beam that represents the resolution limits of the instrument. Moreover, the last residual is added to the final image. This step is of direct meaning for the regularizing property of CLEAN: how does CLEAN fit the gaps in the uv-coverage? Again we assume the model of point-sources from Eq. \eqref{eq: point-sources}. Let us assume that CLEAN has computed a guess model: $M = \sum_l \hat{a}_l \delta_l$, where the weights $\hat{a}_l$ should approximate the true weights $a_l$ sufficiently well. Then, the final residual $R$ reads:
\begin{align}
    R = I^D - B^D * M = B^D * \left( \sum_l (a_l-\hat{a}_l) \delta_l \right), \label{eq: clean_res}
\end{align}
and the final model:
\begin{align}
    I^M = M + R = B^D * \sum_l a_l \delta_l + (\mathds{1}-B^D) * \sum_l \hat{a}_l \delta_l, \label{eq: clean_model}
\end{align}
where $\mathds{1}$ denotes the identity operator. The sum is decomposed in a part that corresponds to the measured Fourier coefficients (first term, convolution with dirty beam sets the Fourier coefficients in the gaps exactly to zero), and a part that corresponds to the uncovered gaps in the uv-coverage (second term, convolution with an `invisible' beam $Id-B$ that is exactly zero for the measured Fourier coefficients and unequal to zero in the gaps). Hence, the model should always fit the data correctly (first term) in the unphysical, ideal situation of an infinite field of view and uniform weighted data without thermal noise and calibration errors. It becomes obvious that CLEAN (assuming that $\hat{a}_l$ are good approximations to the true weights) interpolates to the uncovered gaps in the uv-coverage by assuming that the same pattern of delta components could be used to describe these signals once they were measured. This, however, is problematic primarily for two reasons: first the uv-coverage of a real VLBI array shows rich radial (e.g. a denser coverage on short baselines) and direction-dependent structural patterns (e.g. highly elliptical uv-tracks for some antenna pairs that give rise to only a few directions in the uv-domain). It is far from obvious that these different regions in the Fourier domain should encode the same feature. It is more likely that the small-scale structure hidden on short baselines and the large scale structure on long baselines show less similarity. A more rigorous multi-scalar (and multi-directional) approach is needed to separate these different structural features and to take the structural pattern of the uv-coverage into account. Secondly, the convergence rates and fitting properties in the presence of thermal noise remain unclear \citep{Schwarz1978}. In practice, the CLEAN model often results in severe overfitting when not stopped early enough. This problem is solved by convolving the final model by the clean beam, i.e. the fits to the usual more-poorly-covered long baselines are suppressed generally. However, this only trades the problem of overfitting for a limited resolution that is challenged by modern state-of-the-art imaging algorithms and for an unphysical separation between the final image (that is used for further analysis, but does not fit the visibilities due to the convolution with the clean beam that causes disparity from the observed visibilities) and a model (that fits the visibilities, but is not useful for image analysis). Again a more rigorous multi-scale approach that improves the separation between gaps and covered regions in the uv-coverage (and suppresses the overfits in former one) is desired. 

The regularization introduced by CLEAN can also be visualized in the image-domain instead of the Fourier domain: here the extrapolation into gaps in the fit translates into suppressing sidelobes in the dirty image. Sidelobes are suppressed as the basis functions (delta-functions) are sidelobe-less and the dirty image and the dirty beam consist of the same sidelobe pattern. Hence, deconvolution suppresses sidelobes by subtracting the sidelobe pattern of the dirty beam from the residual. As we will later discuss, this will be a major difference to our new DoB-CLEAN algorithm.

\subsection{Multi-scale CLEAN/wavelets}
multi-scale-CLEAN (MS-CLEAN) methods have been proposed in the past \citep[][]{Bhatnagar2004, Cornwell2008} to mitigate these problems. In a nutshell, the point-like basis functions from CLEAN are replaced by smooth, positive, extended basis functions that are suitable to represent the image structure. \citet{Bhatnagar2004} used Adaptive Scale Pixels (Asp) which could in principle compress any shape. \citet{Cornwell2008} specified this and used tapered, truncated parabolas, a function with a minor difference to Gaussians (i.e. they have a finite support). In particular, \citet{Cornwell2008} mentions that Gaussians would be possible as well, as long as a very high dynamic range is not desired or image-plane support constraints are required. Our new method is based on the spirit of MS-CLEAN developed in these works. But we fit the image with a completely different wavelet-based dictionary resulting in a different imaging procedure. We will theoretically compare our new algorithm with standard MS-CLEAN approaches in more detail in \ref{ssec: comparison}.

\subsection{Alternative Imaging Approaches}
CLEAN and its variants \citep{Hogbom1974, Clark1980, Schwab1984, Bhatnagar2004, Cornwell2008, Rau2011} have been the standard method in VLBI for the last decades. They still remain in use due to their practical nature that allows the astronomer to interact with the imaging manually, to manipulate the data set and to self-calibrate the data set during imaging. We therefore aim to keep this workflow for our new proposed procedure. However, we like to mention the many modern methods developed for VLBI. This includes regularized maximum likelihood (RML) methods \citep[e.g.][]{Carrillo2012, Garsden2015, Akiyama2017, Chael2018, Mueller2022} as well as Bayesian reconstructions \citep[e.g.][]{Junklewitz2016, Cai2018a, Cai2018b, Arras2019, Broderick2020, Broderick2020b, Arras2020}. In comparison to CLEAN the problem is solved by forward modeling instead of inverse modeling.

\section{Algorithm} \label{sec: algorithm}

\subsection{Overview} \label{ssec: overview}
We demonstrated in \citet{Mueller2022} how a multi-scale approach can improve imaging performance. Our algorithm was based on a wavelet-based sparsity promoting (compressed sensing) approach in the RML fashion. In this paper we are interested in a more CLEAN-like algorithm as this working procedure is well established within the VLBI community. In particular, we are proposing a new version of MS-CLEAN \citep{Cornwell2008}, but for the first time we select the basis functions in a way that they fit to the uv-coverage. This provides an optimal selection between observed image features and sidelobes induced by uv-coverage defects. 

We model the true image by a set of extended basis functions (a dictionary) $\Psi = \{ \Phi_0, \Phi_1, ...\}$ instead of delta functions, i.e. $I = \Psi x$ with some coefficient array $x$. We try to recover the coefficient array $x$ from the data and infer the recovered image from there by applying the dictionary on $x$ once more, the recovered image will be $I = \Psi x$ (where $x$ is the recovered array of coefficients). The basis functions $\Phi_i$ have some connection to the Fourier domain: convolving with $\Phi_i$ in the image domain is equivalent to multiplying with the Fourier transform $\mathcal{F} \Phi_i$ in the Fourier domain. The basis functions of the dictionary therefore define filters in the Fourier domain which allow for inserting information of the uv-coverage during the imaging procedure, i.e. every basis function $\Phi_i$ compresses features of a specific set of baselines.

What basis functions are most efficient in that regard? For the purpose of representing the image best, we desire basis functions that are smooth, sidelobe-free and positive \citep[compare the selection of basis functions in][]{Cornwell2008}. For the purpose of fitting the uv-coverage best, basis functions that provide steep radial masks in the Fourier domain and that are optimally orthogonal on each other are desired. These are contradicting requirements. Typical orthogonal wavelet functions (such as Daubechies-wavelets) contain wide sidelobes themselves \citep{Starck2015}. Therefore, we are dealing with two different dictionaries: with a dictionary of (radially) orthogonal wavelets $\Psi^\mathrm{DoB}$, called Difference-of-Bessel (DoB) in the following, that enables the best handling of masks in the Fourier domain and with a dictionary of smooth and clean wavelets $\Psi^\mathrm{DoG}$ that can be used best to describe image features, called Difference-of-Gaussian (DoG) in the following. The two wavelet dictionaries are related to each other such that latter one (the image-driven dictionary $\Psi^\mathrm{DoG}$) contains only the central peak (without sidelobes) of the wavelets of former one (the Fourier domain driven dictionary $\Psi^\mathrm{DoB}$). This is a similar approximation to the one within CLEAN and MS-CLEAN by the transition from the dirty beam to the clean beam, i.e. by fitting a central Gaussian component to dirty beam pattern.

The CLEANing procedure is done with $\Psi^\mathrm{DoB}$. We represent the dirty image by $I^D = B^D * (\Psi^\mathrm{DoB} x)$ and recover iteratively the coefficient array $x$ by CLEAN loops, i.e. we iteratively search for the maximum peak, store this in a list of multi-scalar components and update the residual. The list of multi-scalar components for the final image however is convolved with $\Psi^\mathrm{DoG}$ instead of $\Psi^\mathrm{DoB}$. In this sense, representing the model by shifted and rescaled DoB-wavelets does not suppress sidelobes in the image (since the basis functions $\Psi^\mathrm{DoB}$ contain sidelobes on their own), but works as a feature-finder algorithm that decomposes the dirty image into a list of (extended) multi-scalar basis functions. These are then replaced by more regular basis functions that compress the same image features (the same scales), but suppress the long elongating sidelobes. This is done in an alternating iterative procedure with iterative updates of the residual map: we represent the dirty image by the dictionary $\Psi^\mathrm{DoB}$ by CLEAN loops, we compute a guess solution by replacing the dictionary $\Psi^\mathrm{DoB}$ with the dictionary $\Psi^\mathrm{DoG}$, we update the residual image and repeat these steps until the residual image is noise-like. Opposed to CLEAN, the suppression of sidelobes is not done by finding the CLEAN components and subtracting the dirty beam from the image, but by replacing $\Psi^\mathrm{DoB}$ with $\Psi^\mathrm{DoG}$.

\begin{figure*}
    \centering
    \includegraphics[width=\textwidth]{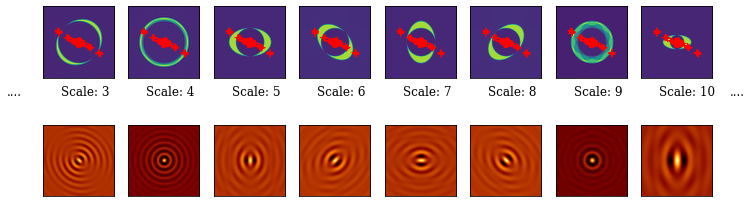}
    \caption{Upper panel: Fourier transform of the used wavelet scales $\Phi_\mathrm{DoB}$ fitted to a synthetic RadioAstron uv-coverage (red points). Shown are the scales of various radial widths (scales 3,4, scales 5-9 and scale 10) and four different elliptical directions. The scale fit to the uv-coverage as they are sensitive to gaps or covered coefficients respectively. Lower panels: The respective wavelet basis function in image domain.}
    \label{fig: scales}
\end{figure*}

In our former paper \citet{Mueller2022}, we presented a novel wavelet dictionary based on the difference of Gaussian method (DoG) that proved flexible enough to compress information about the uv-coverage of the observation. We therefore reuse this dictionary for the image domain $\Psi^\mathrm{DoG}$. It is the canonical extension to orthogonal wavelets to replace the Gaussians in the construction of the DoG-wavelets by modified Bessel-functions of the same width (i.e. the central peak of the Bessel functions has the same width as the Gaussians). The Fourier transform of modified Bessel functions is a uniform disk, hence the Fourier transforms of difference of Bessel (DoB) wavelets are uniform rings. These have non-overlapping support in the Fourier domain, hence are orthogonal. We therefore construct the wavelets for fitting the uv-coverage $\Psi^\mathrm{DoB}$ out of DoB-wavelets. Moreover, we present how to extend this concept also to direction dependent wavelets. Some examples of our sequence of wavelets and their corresponding filter in Fourier domain are presented in Fig. \ref{fig: scales}. Moreover, we present the cross-section of two example wavelets in Fig. \ref{fig: scale_5} and Fig. \ref{fig: scale_7} demonstrating the correspondence between DoB-wavelet scales and DoG wavelets. We present more details on this in the subsequent subsections.

\subsection{Wavelet-basis Functions}
We explain in this section the design of the wavelet functions used in this work. As discussed in Sec. \ref{ssec: overview}, we aim to find a suitable dictionary $\Psi^{DoG}$ that is flexible in its radial scales and smooth to compress image features best, and a dictionary $\Psi^{DoB}$ that corresponds to the same radial (and angular) scales and provides optimal analysis masks in the uv-domain. Our wavelet dictionaries are based on the design of difference of Gaussian (DoG) wavelets that we successfully applied to VLBI imaging in \citet{Mueller2022}. We first summarize the construction of the DoG-wavelet dictionary from \citet{Mueller2022}, before we discuss the straightforward extensions to difference of Bessel functions (DoB) and angular wavelet dictionaries. For more details we refer to \citet{Mueller2022}.
 
One of the most frequently applied continuous wavelet functions is the `Mexican-hat'-wavelet \citep{Murenzi1989, Starck2015} which is known to offer image compressions for a wide range of model images. The `Mexican-hat' wavelet is effectively a (rescaled) Laplacian of Gaussian \citep{Gonzalez2006}. Hence, it is well approximated by the corresponding differential quotient for small variations \citep{Assirati2014}, which we call DoG-wavelet in the following:
\begin{align}
    \Phi_\mathrm{DoG}^{\sigma_1, \sigma_2}(x, y) &= \frac{1}{2 \pi \sigma_1^2} \exp \left( \frac{-r(x,y)^2}{2 \sigma_1^2} \right) - \frac{1}{2 \pi \sigma_2^2} \exp \left( \frac{-r(x,y)^2}{2 \sigma_2^2} \right) \\
    &= G_{\sigma_1} - G_{\sigma_2}, \label{eq: dog}
\end{align}
where necessarily $\sigma_1 \leq \sigma_2$ and $G_{\sigma_j}$ denotes a Gaussian with standard deviation $\sigma_j$. 

In the past, discrete \`{a}-trous wavelet decompositions were of special interest in radio astronomy \citep{Starck2006, Starck2015, pyBDSF2015, Mertens2015, Line2020}. These  wavelet decompositions (called starlet) were successfully applied to imaging and image segmentation. A starlet decomposition can be computed quickly by a hierarchical upstream filtering instead of repeated convolutions in high dimensions. The image is iteratively convolved with a small filter which has typically a small support of only a couple of coefficients. The filter is applied on the output of the preceding filtering operation respectively. In this way a sequence of smoothed images is computed that we denote following our notation in \citet{Mueller2022} by $c_j$, where $j \in [0, 1, 2, ..., J]$ labels the scale. Thus, the scales $c_j$ are smoothed copies of the original (full resolution) image smoothed by $2^j \rho$ pixels, where $\rho$ is the limiting resolution of filter kernel in units of pixels. Wavelets are computed by the difference method:
\begin{align}
\omega_j = c_j - c_{j+1}, \label{eq: omega_c-c}
\end{align}
such that each wavelet scale $\omega_j$ compresses the image information on spatial scales between $2^j \rho$ pixels and $2^{j+1} \rho$ pixels. The sequence of discrete \`{a}-trous wavelets is completed by the final smoothing scale $c_J$. The set $[\omega_0, \omega_1, \omega_2, ..., \omega_{J-1}, c_J]$ is an over-complete representation of the original information, i.e. no information is lost or suppressed during convolution. In particular the image at limiting resolution $c_0$ is recovered by all scales by an easy superposition:
\begin{align}
c_0 = \sum_{j=0}^{J-1} \omega_j + c_J. \label{eq: completeness}
\end{align}
This property proved to be key to our application of wavelets in \citet{Mueller2022}. 

While discrete \`{a}-trous wavelets are very successful in the compression of image information, they are less flexible than a continuous wavelet transform due to the inherent upsampling by a factor of two. Hence, they lack the ability to fit sufficiently to the more complex uv-coverage of real VLBI arrays. Therefore, we define a flexible wavelet dictionary out of DoG-wavelets in the same procedure as was done for the \`{a}-trous wavelet: We define an increasing set of widths $[\sigma_0, \sigma_1, \sigma_2, ..., \sigma_J]$ and compute the filtered scales of the original image by convolving with Gaussians, i.e. $c_j = G_{\sigma_j} * I$ (where I denotes the original image). It is (compare \citet{Mueller2022}):
\begin{align}
\omega_j = c_j - c_{j+1} = \Phi_\mathrm{DoG}^{\sigma_j, \sigma_{j+1}} * I,
\end{align}
and the complete set of scale satisfies the completeness relation Eq. \eqref{eq: completeness} again. If the original image $I$ is noisy, the scales $\omega_j$ will be noisy as well with a scale-specific noise-level. All in all, the DoG-wavelet decomposition operation reads:
\begin{align}
\Psi^\mathrm{DoG}: I \mapsto [\Phi_\mathrm{DoG}^{\sigma_0, \sigma_{1}} * I, \Phi_\mathrm{DoG}^{\sigma_1, \sigma_{2}} * I, ..., G_{\sigma_J} * I].
\end{align}

\begin{figure}
    \centering
    \includegraphics[width=0.5\textwidth]{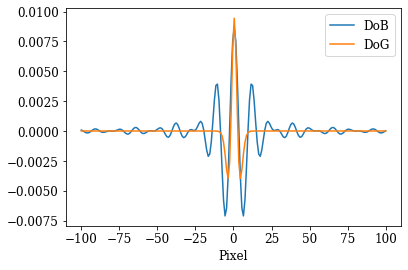}
    \caption{Cross-section of the DoB and DoG wavelet scale 5 presented in Fig. \ref{fig: scales}. The DoB-wavelet fits the central peak, but suppresses the extended sidelobes.}
    \label{fig: scale_5}
\end{figure}

\begin{figure}
    \centering
    \includegraphics[width=0.5\textwidth]{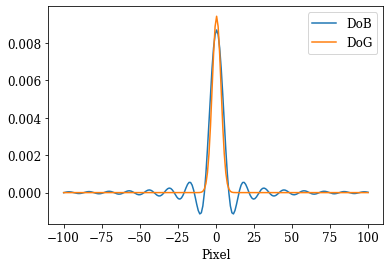}
    \caption{Cross-section of the DoB and DoG wavelet scale 7 presented in Fig. \ref{fig: scales}. The DoB-wavelet fits the central peak, but suppresses the extended sidelobes.}
    \label{fig: scale_7}
\end{figure}

Convolutions in the image domain translate to multiplicative masks in the Fourier domain. The Fourier transform of a DoG-wavelet is a difference of non-normalized Gaussian functions:
\begin{align}
    \mathcal{F} \Phi_\mathrm{DoG}^{\sigma_j, \sigma_{j+1}}(u,v) \propto \exp \left( -2 \pi^2 \sigma_j^2 q(u,v)^2  \right) - \exp \left( -2 \pi^2 \sigma_{j+1}^2 q(u,v)^2  \right), \label{eq: fourier_dog}
\end{align}
which defines ring-like masks in the uv-domain, compare \citet{Mueller2022}.  To have steep and orthogonal masks however, we propose to replace Gaussians in the construction of wavelets by spherical Bessel functions. Hence:
\begin{align} \nonumber
    &\Phi_\mathrm{DoB}^{\tilde{\sigma}_j, \tilde{\sigma}_{j+1}} (x, y) = \\
    & \frac{1}{\tilde{\sigma}_j r(x, y)} J_1(2 \pi r(x, y) / \tilde{\sigma}_j) - \frac{1}{\tilde{\sigma}_{j+1} r(x, y)} J_1(2 \pi r(x, y) / \tilde{\sigma}_{j+1}),
\end{align}
where $J_0$ denotes the Bessel function of first order and $\tilde{\sigma}_j$ the widths of the Bessel functions. The widths for the DoB-wavelets $\tilde{\sigma}$ are typically not the same as the widths that we use for the DoG-wavelets. In fact, we will determine $\tilde{\sigma}_j$ first by fitting DoB-wavelets to the uv-coverage as described in Sec. \ref{sec: widths}, then we will select the widths for the DoG-wavelets $\sigma_j$, such that the correlation between DoB-wavelet and DoG-wavelet is maximal, see our demonstration in Fig. \ref{fig: scale_5} and Fig. \ref{fig: scale_7}.

It is in two dimensions:
\begin{align}
    \mathcal{F}^{-1} (1_K)(r) = \frac{K}{r} J_0(2 \pi K r) =: \tilde{J}_{1/K} (r),
\end{align}
where $1_K$ is a disc with radius $K$ in Fourier domain. Hence the Fourier transform of the DoB-wavelet is a ring-shaped mask with step-like cross-section:
\begin{align}
    \mathcal{F} (\Phi_\mathrm{DoB}^{\tilde{\sigma}_j, \tilde{\sigma}_{j+1}}) (k) = 1_{1 / \tilde{\sigma}_j} (k) - 1_{1 / \tilde{\sigma}_{j+1}} (k).
\end{align}
All DoB-wavelets are therefore orthogonal to each other as the Fourier transform is a unitary operation and the wavelets $\Phi_\mathrm{DoB}^{\tilde{\sigma}_j, \tilde{\sigma}_{j+1}}$ have non-overlapping support in Fourier domain.

Up until now we have discussed only the case of radially symmetric wavelets. To match the patterns in uv-coverages of real VLBI arrays, a direction-dependent dictionary is desired as well. This extension is straightforward by replacing the radial symmetric Gaussian/Bessel functions by elliptical beams. We now demonstrate the construction of direction-dependent wavelet dictionary for the DoG-wavelets. The construction for DoB-wavelets is analogous.

We start with radial widths $[\sigma_j]$, and $N$ angles $\alpha_0, \alpha_1 = \alpha_0 + \frac{2 \pi}{N}, ...,  \alpha_{N-1} = \alpha_0 + \frac{2 \pi}{N} (N-1)$ equidistantly distributed on a circle. We then calculate radial Gaussians $G^r_{\sigma_j}$ and elliptical Gaussians $G^e_{\sigma_j, \sigma_{j+1}, \alpha_i}$ with major axis $\sigma_{j+1}$ and minor axis $\sigma_{j}$ rotated by an angle $\alpha_i$. Hence, when decomposing an image $I$ we compute filtered smoothed, radial bands $c^r_j = G^r_{\sigma_j} * I$ and elliptical bands $c^e_{j,i} = G^e_{\sigma_j, \sigma_{j+1}, \alpha_i} * I$ and compute wavelets by:
\begin{align}
    \omega_{j,i} = c^r_j - c^e_{j, i}. \label{eq: omega_radial_elliptic}
\end{align}
Due to the combination of radial wavelets and elliptic wavelets $\omega_{j,i}$ has a single directionality which is necessary to capture the direction dependence. Moreover, a construction in the spirit of Eq. \eqref{eq: omega_radial_elliptic} allows to complete the dictionary easily, i.e. to satisfy a completeness property similar to Eq. \eqref{eq: completeness}. We complete the set of wavelets with the residual scales $\omega_{j, N} = \frac{1}{B_j} \sum_{i=0}^{N-1} c^e_{j, i} - c^r_{j+1}$ (where $B$ is a normalization constant such that $\norm{\omega_{j,N}} = \norm{\omega_{j, N-1}}$ for a response to a delta source). The final smoothing scale $\omega_{J} = c^r_J$. We present the complete action of the dictionaries $\Psi^\mathrm{DoG}$ and $\Psi^\mathrm{DoB}$ in Appendix \ref{app: dictionaries}. The complete set of wavelet scales $\{ \omega_{j,i}, \omega_J \}$ satisfies a completeness property again:
\begin{align}
    N c^r_0 = \sum_{j=0}^{J-1} \left( \sum_{i=0}^{N-1} \omega_{j,i} + B_j \omega_{j, N} \right)+ N \omega_{J}.
\end{align}

\subsection{Radial Widths} \label{sec: widths}
We explain in this subsection which widths $\sigma_0 < \sigma_1 < ... < \sigma_J$ are selected to get an optimal fit to the uv-coverage. The selection of these basis functions has to be done prior to the imaging procedure. The basis functions are selected in a way that they allow for an optimal separation between covered Fourier coefficients and unsampled Fourier coefficients, such that some wavelet basis functions compress Fourier information that is covered by data and the remaining one compress scalar information that has not been observed (gaps). The only important criterion here is whether a scale is sampled or not. For the selection of scales we do not process the signal strength or phase observed in the visibilities. Hence, at this stage only the uv-coverage is processed. During the imaging a least-square fit to the visibilities at every scale will be done, with effective suppression of the non-covered scales.

This selection is similar to the procedure that we already proposed in \citet{Mueller2022}. We are selecting the radial widths only, the angular elliptical beams are always constructed from the same array of angles equidistantly distributed on a circle at all radial scales $\sigma_j$.

The angle-offset $\alpha_0$ is chosen to be the rotation of the major axis in the clean beam. For the selection of the radial scales, we extract the array of uv-distances, sort this array in increasing order and look for jumps in the sorted new array. If the increase from one component in the sorted array to the next one exceeds some (manually chosen) threshold, we store the radial baseline lengths $q(u, v)$ for the two neighboring data points in a list of lengths in the Fourier domain. We translate these lengths in the Fourier domain into an increasing list of radial widths of spherical Bessel functions in the image plane $[\sigma_i]$ by inverting. Finally we complete this list: if there is an index $i$, such that $2 \sigma_i < \sigma_{i+1}$, we add a scalar width $\sigma = (\sigma_{i+1}+\sigma_i)/2$ to avoid to large gaps between consecutive widths.

The resulting DoB-wavelet dictionary fits well to the uv-coverage, compare the Fourier filters presented in Fig. \ref{fig: scales}. As a next step, we have to find the radial widths for the DoG-wavelets. Recall that the DoB-wavelets were constructed with the \`{a}-trous differential method. We construct the DoG-wavelets in the same way. We therefore fit Gaussians with varying radial widths to the central peaks of the spherical Bessel functions of widths $[\sigma_i]$. We then construct the DoG-wavelets by the differential method from these Gaussians. The resulting DoG-wavelets are approximating the central peaks of the DoB-wavelets, but without the wide sidelobes of the DoB-wavelets. This is demonstrated in Fig. \ref{fig: scale_5} and Fig. \ref{fig: scale_7}. A sequence of examples of selected DoB-scales and the respective Fourier transform masks is shown in Fig. \ref{fig: scales}.

The threshold parameter used in this procedure to identify the gaps in the uv-coverage is a free parameter. If it is chosen too large, smaller gaps will be skipped. If it is chosen too small, the number of selected basis functions increases and samples the uv-coverage more accurate than might be necessary. In this work the threshold was always chosen such that the most obvious radial gaps are kept and the number of basis functions does not exceed fifty to assure good numerical performance, but this may vary based on the array configuration.

\subsection{Scale-selection Criterion} \label{ssec: selection}
Let us assume first orthogonal wavelet functions $\Phi_j$, where $j$ counts the scale.

Let us assume the true image $I$ is modeled by a sum of wavelets:
\begin{align}
I = \sum_{j, n, l} a_{j,n,l} \Phi_{j,n} * \delta_l,
\end{align}
where $j$ labels the (radial) scale in use, $n$ labels the angle of the ellipse and $l$ labels the pixels in the image (position of the wavelet). This assumption is well motivated by the great success that wavelet-based segmentation, image compression and decomposition have in radio astronomy \citep{Starck2015, Mertens2015, Line2020} and in particular better motivated than the implicit pixel-based CLEAN assumption. Note, that if we replace one scale $\Phi_j$ by two smaller scales $\Phi_{j1}$ and $\Phi_{j2}$ satisfying $\Phi_j = 2 \Phi_{j1} = 2 \Phi_{j2}$, it would hold $a_j = a_{j1} = a_{j2}$. Hence, the magnitude of $a_{j,n,l}$ does not depend on the relative size of the corresponding wavelet. Thus, in every CLEAN iteration we would like to find the biggest $a_{j,n,l}$ still in the dirty image. However, some scales are not covered in the data. We therefore update our goal: we want to find the biggest $a_{j,n,l}$ still in the residual \textit{for which the corresponding wavelet basis function $\Phi^\mathrm{DoB}_{j,n}$ corresponds to sampled Fourier coefficients}. How much a scale is covered in the data is measured by the dirty beam: if one scale is covered (i.e. the Fourier coefficients compressed by this scale are sampled), the product $\norm{ \mathcal{F}\Phi^\mathrm{DoB}_{j,n} \cdot S} = \norm{ \Phi^\mathrm{DoB}_{j,n} * B^D}$ is large and vice versa (where $S = \mathcal{F} B^D$ is a pixel-based mask in the Fourier domain). We therefore formulate our selection criterion as follows: we want to find the scale $j$, angle $n$ and the position $l$, such that:
\begin{align}
\{j,n,l\} \in argmax \frac{\norm{\Phi_{j, n} * B^D}}{\norm{\Phi_{j, n}}} a_{j,n,l}
\end{align}
is maximal, where $B^D$ denotes the dirty beam. The question on hand is, how could we fulfill this criterion in the selection of peaks. Note that the model parameters $a_{j,n,l}$ are not known to us. In fact, we want to determine them from the dirty image (in the following labeled by $I^D$).

We will demonstrate that we fulfill our criterion if we convolve the dirty image with the beam:
\begin{align}
B^\phi =  \frac{1}{\norm{\Phi_{i, m} * B^D} \norm{\Phi_{i, m}} } \Phi_{i,m} * B^D \label{eq: selection}
\end{align}
and search for the maximum over the scales $i$, the angle $m$ and the position of the peak $k$, i.e. $\{i_\mathrm{max},m_\mathrm{max},k_\mathrm{max}\} \approx \{j,n,l\}$. In fact, when we search for the peak over all these scales we solve the optimization problem:
\begin{align}
\{i,m,k\} \in argmax_{i,m,k} \frac{1}{\norm{\Phi_{i,m} * B^D} \norm{\Phi_{i,m}} } \left( \Phi_{i,m} * B^D * I^D \right) (k). \label{eq: argmax}
\end{align}
A detailed proof of this identity, i.e. that we match our selection criterion Eq. \eqref{eq: selection} with the optimization strategy Eq. \eqref{eq: argmax}, is presented in App. \ref{app: proof}.

\subsection{Pseudocode/Implementation}

We summarize DoB-CLEAN in Tab. \ref{alg: wclean}. First we compute the dirty image $I^D$ and the dirty beam $B^D$ as usual for CLEAN. We then fit the scale widths $\{ \tilde{\sigma}_i \}$ to the uv-coverage in the way described in Sec. \ref{sec: widths}. Out of these scale-widths $\{ \tilde{\sigma}_i \}$ we construct the DoB-wavelet dictionary $\Psi^\mathrm{DoB}_\mathrm{clean}$ by the difference method from modified Bessel functions. We find the widths of the corresponding DoG-wavelet dictionary by fitting the central peak of the modified Bessel functions with Gaussian functions. We define the DoG-wavelet dictionary $\Psi^\mathrm{DoG}_\mathrm{clean}$ by the difference method again from these Gaussians.

Recall from Sec. \ref{ssec: selection} that for the weights of the different scales and for the selection of the correct scales, the convolution of our wavelet-functions with the dirty beam plays a vital role, i.e. compare Eq. \eqref{eq: argmax}. We therefore absorb the dirty beams in the  definition of the dictionaries to reduce computational cost, i.e. we compute a `dirty' DoB-wavelet dictionary: $\Psi^\mathrm{DoB}_\mathrm{dirty} = D * \Psi^\mathrm{DoB}_\mathrm{clean}$.

Now, before the cleaning process starts, we can precompute the data products required for the cleaning iterations later on. We decompose the dirty image by $\Psi^\mathrm{DoB}_\mathrm{dirty}$ for the multi-scalar search of the maximal peak in the residual during the minor loop. We have to use the `dirty' dictionary here according to our scale-selection criterion Eq. \eqref{eq: argmax}. Moreover, we have to decompose the dirty beam by our set of basis function that will represent the image in the first instance, i.e. by $\Psi^\mathrm{DoB}_\mathrm{clean}$. These scales of the dirty beam $B^D_j$ will be subtracted from the residual during the minor loop of the CLEAN iterations. It is further beneficial to compute the subtraction from the image-scales $I_i$ scale-by-scale independently instead of subtracting the complete beam $B^D_j$ from the residual and recomputing the image-scale decomposition newly every iteration. Hence, we precompute the scalar decomposition of the beam-scales $B^D_j$ by the `dirty' dictionary $\Psi^\mathrm{DoB}_\mathrm{dirty}$ as well. Moreover, we normalize these beams by their maximal peak. Note that these data products ($B^D_{i,j}$) have to be computed only once before the CLEAN loops start until the dirty beam is changed (due to a new weighting scheme, flagging of data, and other operations). Later on, only convolutions of these wavelets with delta-components have to be computed. Hence, we can compute the subtractions of the multi-scalar beams very efficiently by shifting and rescaling the precomputed beam-scales $D_{i,j}$. Finally, we precompute the multi-scalar weights $w_j$ that we explained in Sec. \ref{ssec: selection}, i.e. the denominator in Eq. \eqref{eq: argmax}. 

As outlined before, we carry out the CLEANing procedure by iterating between a CLEAN loop (with DoB-wavelets as basis functions, inner loop) and switching between dictionaries (from DoB-dictionary to DoG-dictionary, outer loop). In the inner loop we iteratively search for the largest peak among the image scales and store the position, the scale and intensity in a list of delta components. We then update the residual scale-by-scale by subtracting the recently found component. After a sufficient number of iterations, we compute a model $M$ by summing our stored delta components, but applying the dictionary $\Psi^\mathrm{DoG}_\mathrm{clean}$ instead of the dictionary $\Psi^\mathrm{DoB}_\mathrm{clean}$ (outer loop). After this switch of dictionaries we have to reinitialize the residual and the residual-scales for the next DoB-CLEAN runs. At this step also further data manipulation steps, such as flagging, self-calibration, thresholding the image or projecting to positive fluxes, could be applied as required depending on the data set under consideration. We also refer to Fig. \ref{fig: sketch_dobclean} in which we demonstrate the working procedure of DoB-CLEAN on one of the synthetic data sets that will be used in Sec. \ref{sec: synthetic_test}. The dirty beam is successfully cleaned out of the image by the representation by DoB-wavelets (small residual). However, the wavelets itself contain sidelobes and hence the DoB model has these sidelobes as well. By switching to DoG wavelets we get a physical and smooth model that still fits the visibilities.

\begin{table}
\caption{Wavelet CLEAN Algorithm.}

\begin{tabular}{p{0.45\textwidth}}
\hline \\
\end{tabular}

\begin{algorithmic}
\Require Dirty Image: $I^D$
\Require Dirty Beam: $B^D$
\Require gain: $g$
\Require scale-widths for Wavelet-decomposition (DoB): $\{ \tilde{\sigma}_j \}$ fitted to uv-coverage
\\
\State Define `clean' DWT by difference of Bessel functions with scales $\tilde{\sigma}_j$: $\Psi^{DoB}_{clean}$
\State Fit Gaussian functions to the central peaks of the Bessel functions, define a difference of Gaussians (DoG) dictionary by these fits: $\Psi^{DoG}_{clean}$. Note that this dictionary approximates the Bessel dictionary, but without the sidelobes.
\\
\State Define `dirty' DWT by DoB with `dirty' scales $\hat{\sigma}_j$: $\Psi_{dirty}$, where $\Psi^{DoB}_{dirty} = \Psi^{DoB}_{clean} * B^D$  
\\
\State Decompose dirty image by $\Psi^{DoB}_{dirty}$: $I^D = \sum I^D_j$
\State Decompose dirty beam by $\Psi^{DoB}_{clean}$: $B^D = \sum B^D_j$
\State Decompose the scales of the dirty beam by $\Psi^{DoB}_{dirty}$: $B^D_j = \sum B^D_{ij}$
\\
\State Find normalization constants: $n_j = max(B^D_{j})$
\State Normalize beam and psf by $n_j$: $B^D_{ij} = B^D_{ij}/n_j$ ... for all $i$ and $j$
\\
\State Find weights: $w_j = \frac{1}{\norm{\Psi^{DoB}_{dirty}} \cdot \norm{\Psi^{DoB}_{clean}}}$ (these weights were proven to be optimal)\\
\\

Initialize restoring image: $M = 0$
\While{residual not noise-like}

\While{number of maximal iterations not reached}
\State Find Maximum of $[w_j \cdot abs(I_j)]$ searching over scales $j$ and pixels $k$
\State Store maximum $I_j^k \cdot \delta_j^k$ in list of components
\State For every scale $l$: $I_l = I_l - g \cdot I_j^k \cdot shift(B^D_{lj}, k)$   
\EndWhile \\
\State $M = M + \sum g \cdot I_j^k \cdot \Psi_{clean}^{DoG} \delta_j^k$ 
\State Update dirty image/residual: $I^D = I^D - B^D * M$
\State Reinitialize the decomposition: $\Psi^{DoB}_{dirty}$: $I^D = \sum I^D_j$
\State optional: self-calibration 
\State optional: project solution to positive values
\EndWhile \\

\State Add residual image: $M = M + \sum I^D_j$
\Ensure $M$ is approximation to true sky image
\end{algorithmic}

\begin{tabular}{p{0.45\textwidth}}
\hline \\
\end{tabular}

\label{alg: wclean}
\end{table}

\subsection{Comparison to CLEAN and MS-CLEAN} \label{ssec: comparison}
DoB-CLEAN succeeds over CLEAN by using a multi-resolution approach to imaging. This allows for a better separation between image features and sidelobes. Hence, DoB-CLEAN provides more reasonable regularization. Let us repeat the regularization analysis presented in Eq. \eqref{eq: clean_res}-\eqref{eq: clean_model}. We assume that the true model reads as:
\begin{align}
    I = \sum_l a_l \Psi_l^\mathrm{DoB}.
\end{align}
Note that although the wavelet functions $\Psi_l^\mathrm{DoB}$ contain clearly unphysical sidelobe structures, this is not a stronger assumption than the point source assumption that we did for the analysis of CLEAN, i.e. Eq. \eqref{eq: dirty_map}, due to the completeness of the wavelet dictionary Eq. \eqref{eq: completeness}. The dirty image is then:
\begin{align}
    I^D = \sum_l a_l \Psi_l^\mathrm{DoB} * B^D \approx \sum_i a_i \Psi_i^\mathrm{DoB} * B^D,
\end{align}
where the indices $i$ are a typically sparse subset of the space of indices $l$. This harvests one of the main advantages of DoB-CLEAN over CLEAN. While the sparsity assumption that is hard-coded in CLEAN is somewhat dubious, in particular if extended structures are studied, DoB-CLEAN tries to sparsely represent the dirty image with a dictionary especially designed for this purpose. The wavelet functions that correspond to scales in the Fourier domain that are not covered can be omitted in the sum above (the convolution with the dirty beam vanishes) and the sparsity assumption is really fulfilled. The dirty image is modeled by:
\begin{align}
    M^D = \sum_i \hat{a}_i \Psi^\mathrm{DoB},
\end{align}
where $\hat{a}_i$ denotes the estimated approximations to the true coefficients $a_i$ calculated by DoB-CLEAN. The cleaned image model reads:
\begin{align}
    M = \sum_i \hat{a}_i \Psi^\mathrm{DoG}.
\end{align}
Hence, the residual is:
\begin{align}
    R = I^D - B^D * M \approx B^D * \sum_i \left[ a_i \Psi_i^\mathrm{DoB} - \hat{a}_i \Psi_i^\mathrm{DoG} \right].
\end{align}
Thus: 
\begin{align}
    I^M = M+R =  B^D * \sum_l a_l \Psi_l^\mathrm{DoB} + (\mathds{1} - B^D) * \sum_i \hat{a}_i \Psi^\mathrm{DoG}.
\end{align}
Again we recover the correct data fit for the covered scales. In the second term we process information from covered scales only (indices $i$). We therefore extrapolate the data fit to the gaps in the uv-coverage by the same core-information as the signal from the covered scales (the DoG-wavelets fit the central peak of the DoB-wavelets), but we suppress the sidelobes. This can be translated to the Fourier domain in that we copy the same information that we recovered from covered scales also in uncovered scales, but the importance decreases with distance from the covered Fourier coefficients. We therefore, in contrast to CLEAN, recover the final model from the measured visibility points only and suppress the information in the gaps to a level, such that the final recovered model appears smooth and free of sidelobes, but no image features are hidden in the gaps. This seems to be an optimal criterion for us given the sparsity of the measured visibilities. We will expand more on how CLEAN and DoB-CLEAN fit the gaps in the uv-coverage in Sec. \ref{ssec: artifacts}.

The replacing of DoB wavelets by DoG wavelets is similar to a multiscalar variant of replacing the dirty beam by the clean beam as done for CLEAN. However, there are subtle differences. For DoB-CLEAN the convolution is not done as a final step, but takes place within the minor loop, such that the new residuals are computed after convolution with $\Psi^\mathrm{DoG}$. Moreover, compare Tab. \ref{alg: wclean}, we replace in the minor loop the `dirty' scales $\Psi^{DoB}_{dirty}=\Psi^{DoB}_{dirty}*B^D$ with the `clean' scales $\Psi^{DoG}_{clean}$. Since the basis functions are already extended and fit to the uv-coverage, in particular to the limiting resolution, a final additional convolution with a clean beam is not needed. This convolution is unphysical as it introduces a disparity between the model fitted to the visibilities and the final image. Our algorithm directly computes a clean (i.e. free of sidelobes) model that fits to the visibilities and that matches our perception of astronomical reality, i.e. solves this disparity.

We shall discuss the convergence of DoB-CLEAN shortly at this point. If the model is composed of extended DoG wavelet functions with widths equivalent to the limiting resolution, an additional convolution with the dirty beam to compute the residual could smear out the model image even more and cause divergence. This however is prevented by the scale selection criterion Eq. \eqref{eq: argmax}. Since we convolve the dirty image another time with the dirty beam to find the optimal scale, we select smaller scales (already respecting the fact that another convolution for the computation of the residual will smear out features).

DoB-CLEAN is based on the ideas pioneered in multiresolution CLEAN methods \citep{Bhatnagar2004, Cornwell2008, Rau2011}. However, our new method has some significant differences. Most obviously we use different dictionaries than previous works. MS-CLEAN basis functions are selected on a best effort basis manually \citep{Cornwell2008}. Asp-CLEAN \citep{Bhatnagar2004} is a variant of MS-CLEAN in which the proper scale widths of the basis functions (Asps) are selected by a fit to the data alternating with the minor loop iterations. Asp-CLEAN therefore shares some more philosophical similarities with DoB-CLEAN than standard MS-CLEAN. However, the basic outline remains the same: basis functions are selected based on the image domain to describe the perceived image structure best, thereby solving practical issues of CLEAN in representing extended emission. \citet{Cornwell2008} defined three requirements for such basis functions: each basis function should be astrophysically plausible, they should be radially symmetric and the shape should allow support constraints (although the latter one can be weakened). Opposed to that, our dictionaries are designed on different requirements: we designed wavelet basis functions $\Psi^\mathrm{DoB}$ that fit to the uv-coverage, i.e. that sparsely represent the dirty image. Hence, opposite to MS-CLEAN and Asp-CLEAN, our selection of scales is purely driven by the instrument and no perception of the image structure. This highlights a specific difference to Asp-CLEAN: in Asp-CLEAN the used scales are fitted to optimally fit the observed visibilities in every iteration and this selection strongly affects the minor loop iterations. In DoB-CLEAN, only the uv-coverage, not the visibilities, is used to define scales and the selection of which scales fits the visibilities ideally is controlled by the minor loop. Moreover, we use for the first time a multi-directional dictionary. These requirements are not compatible. This has a couple of consequences that cause DoB-CLEAN to differ from MS-CLEAN algorithms. MS-CLEAN and Asp-CLEAN use the minor and major loops to suppress sidelobes (compare our discussion in \ref{ssec: clean}) by a sparse representation of the true model. DoB-CLEAN uses the minor and major loop of CLEAN to find a sparse representation of the dirty image (not the true image). This makes the use of a second dictionary $\Psi^\mathrm{DoG}$ and a switch between both dictionaries needed. Sidelobes are suppressed by replacing the DoB-wavelets (with large sidelobes) by the DoG-wavelets (without sidelobes). $\Psi^\mathrm{DoB}$ features some more advantages: it is orthogonal in radial dimension. Hence, in DoB-CLEAN scalar features that for example only affect intermediate baselines, but not long or small  baselines can be expressed sparsely while in MS-CLEAN and Asp-CLEAN every basis function necessarily affects the shortest baselines. In particular, there is only one scale $c_J$ that transports flux in the image (compare Eq. \eqref{eq: omega_c-c} and Eq. \eqref{eq: completeness}), all other scales have integral zero. The orthogonality offers the additional advantage that a solid scale-selection criterion could be derived (see Sec. \ref{ssec: selection}), opposite to \citet{Cornwell2008} where the selection of the correct scale is done in an ad-hoc manor by manually choosing a specific scale-bias. We, however, select for the first time the scale that provides the largest correlation to the dirty image. Moreover, the basis function dictionary is complete. Hence, opposite to Asp-CLEAN and MS-CLEAN, there is no doubling of information compressed at different scales.

All in all, compared to CLEAN and MS-CLEAN, DoB-CLEAN succeeds in two important aspects. First, the regularization property (i.e. how to fill the gaps in uv-coverage) is more reasonable. Second, in CLEAN \citep{Hogbom1974} and in MS-CLEAN \citep{Cornwell2008} the final model is blurred with the clean beam, which causes an unphysical separation between model and image as described in the introduction. In DoB-CLEAN however, the basis functions are already extended functions that represent the image features well and are used to fit to the visibilities. Thus, theoretically a final convolution with the clean beam is not needed making the computed image the same as the computed model.

\subsection{Software and Pipeline}

The method has been implemented in the new software package \textit{MrBeam} which makes explicit use of \textit{ehtim} \citep{Chael2018} and \textit{regpy} \citep{regpy}. We designed the user interface to resemble standard
VLBI software packages such as \textit{Difmap} \citep{Shepherd1997}. This has several practical benefits: it resembles the way of working common to scientists. Hence, \textit{MrBeam} allows for the typical tools of interactive manipulation, visualization and inspection of data known from CLEAN softwares: interactive drawing of CLEAN windows (search masks for peaks in the residual), the option for various weighting schemes, taperings and flagging of data, a hybrid self-calibration routine, etc. . This proved practical in the past to address data corruption and calibration issues. However, the practical use of interactive tools remains restricted to small arrays in \textit{MrBeam} as the multiscalar image decompositions have to be recomputed every time the weights or gains have been updated.

In principle DoB-CLEAN needs two stopping rules to be specified. Firstly, we have to specify after how many iterations we want to stop the overall CLEANing procedure (stopping rule 1 in Tab. \ref{alg: wclean}). Secondly, we have to determine for how many iterations do we want to represent the image with DoB-wavelets before we perform the change to the DoG-wavelets (stopping rule 2 in Tab. \ref{alg: wclean}). The former stopping-rule is defined by the noise level of the observation and the current residual. We do not provide a quantitative stopping criterion here but stopped the iterations whenever the residual image looked Gaussian-like and the residuals did not reduce significantly with further iterations. For the latter stopping rule, changing the dictionaries every iteration proved to be the most practical solution, i.e. we update the model image every iteration.

The fitting of the observed visibilities by extended, specially-designed basis functions proved to be helpful in introducing regularization. However, to account for every not-fitted source of flux in the final image, it could be beneficial to clean the already-cleaned residual with several H\"{o}gbom CLEAN iterations on the complete field to improve the fit to observed visibilities. We provide such an option in the software package \textit{imagingbase} underlying this work. However, this finalization step was not found to amend the final model on a level visible by eye.

Lastly, we would like to comment on the use of CLEAN windows. In standard H\"{o}gbom CLEAN windows are essential in the early iterations of the CLEANing and self-calibration to separate the essential true sky brightness distribution from sidelobes. After several iterations the residual is smaller, the sidelobes are suppressed and the underlying image structure becomes visible. The windows can be drawn larger. However, for DoB-CLEAN drawing sophisticated windows did not prove to be essential at all. The sidelobe structure of the beam is imprinted in the basis functions of the DoB-wavelet dictionary and the role of the convolution with the dirty beam is in particular represented, for the first time, in our scale-selection criterion. The maximal correlation is achieved when the multi-scalar component is centered in the sidelobe structure and components are not falsely set in the sidelobes, but rather where the true sky brightness distribution is located. Hence, for our tests on synthetic data in Sec. \ref{sec: synthetic_test} we imaged with DoB-CLEAN on the complete field of view without setting any window.

\subsection{Post-processing} \label{ssec: postprocessing}

The multi-scalar and multi-directional decomposition offers rich possibilities for post-processing. The multi-scale dictionary $\Psi$ provides control over the fit of the model in the gaps within the uv-coverage. This is a great advantage of DoB-CLEAN. In particular, we can identify the image features that are present in the observation and those that are not covered. The signal from the latter is suppressed. In this sense, we construct a mostly sidelobe-free representation of the robustly-measured image information. However, we can use this information as well to reintroduce missing scales in the observation to the image. This step should be done with relative caution as we are adding extrapolated signals.

We implemented and tested the most natural approach to reintroduce missing information in the image, i.e. by interpolating between neighboring scales. For that we first have to identify which scales are labeled as uncovered (i.e. which scales do we have to add to the image in post-processing). We can use the scale-selection criterion here again: we define a threshold $t$ (usually we use $t=0.1$), compute the initial dirty beam with uniform weighting, and label scales as missing if:
\begin{align}
    \frac{ \norm{ \Psi_l^\mathrm{DoB} * B^D} }{ \norm{\Psi_l^\mathrm{DoB} }}< t
\end{align}
For each of these missing scales, we search for the next smaller scale in the same direction (for elliptical scales) and the next larger scale in the same direction and interpolate the coefficient array for the missing scale between these two. We evaluate the performance of post-processing by missing scales in Sec. \ref{ssec: artifacts}. In a nutshell, adding missing (not measured) scales to the image proved useful to suppress artifacts that are introduced by gaps in the uv-coverage. However, this option should be used only with relative caution as signal is predicted for Fourier coefficients that are not constrained by observations, i.e. false image features could be added to the reconstruction when the adding of the missing scales is overdone. While it is a natural choice to interpolate the missing scale from adjacent scales, this does not always have to be the best option. This is in particular true when the structures at various scales have only a small correlation as common for example in VLBI studies of jets powered by an active galactic nuclei (AGN). The bright small-scale features (VLBI-core, innermost jet) and the large scale features (extended jet emission) can vary in morphology, localization and orientation \citep[e.g. compare the multifrequency studies in][with highly varying morphologies between scales]{Kim2020}. Recent progress in multifrequency observations, and the ongoing combination of short baseline and long baseline arrays (and consequently the desire to map galactic structures on a range of spatial scales) may highlight the issue raised above further.

\subsection{Numerical Challenges}
In this subsection we present some numerical issues and challenges for DoB-CLEAN and possible strategies to resolve them.

As the DoB-wavelets are designed to define steep, orthogonal masks in the uv-domain, one has to deal with the Gibbs-phenomenon at the edges of these masks. We found that the field of view should be large enough, such that roughly ten sidelobes of the spherical Bessel functions still fit in it to avoid numerical issues by the Gibbs phenomenon. Additionally, it proved beneficial to fight the rapid accumulation of numerical errors by reinitializing the decomposition of the dirty image from time to time. 

Low-level negative fluxes are introduced into the images by the basis functions itself and have to be negated by neighboring scales, see the completeness relation Eq. \eqref{eq: completeness}). This however also offers a great advantage of DoB-CLEAN over CLEAN. Due to the completeness relation Eq. \eqref{eq: completeness} and the explicit allowance of negative wavelet coefficients, every structure in the current model could in principle be deleted again or completely altered and partly negated by other scales in later iterations. This is more difficult in CLEAN where falsely-set components (e.g. due to corrupted data, calibration issues or falsely-identified windows) are typically removed from the model by flagging manually. Hence, DoB-CLEAN interacts well with extended starting models similar to the working procedure standard in RML methods (iterative imaging with a new starting model and blurring). We therefore have a new, RML-inspired ad-hoc method to avoid negative fluxes in the final image: alternating with imaging we threshold (and blur) the image to the significant flux and reinitialize the residual and the DoB-CLEAN parameters with the thresholded model as a starting model.

After some iterations we project the recovered model to the significant fluxes (i.e. we threshold the model by a small fraction, typically one percent of the peak flux, and in particular project all negative fluxes to zero) and blur the image by the nominal resolution. We take this image as a proper starting model for the next imaging rounds. We recompute the residual and the corresponding decomposition and proceed with the CLEANing with the thresholded model as a starting model. This strategy is well motivated, every high dynamic range image structure that might be falsely deleted from the model, is reintroduced in the newly computed residual and will be reintroduced to the model in the subsequent CLEANing loops. In particular, a worsen resolution after blurring will be corrected later by readding small scale DoG wavelets that shift power from larger scales to smaller scales. As a weaker version of this strategy we also can project only the negative fluxes to zero flux (i.e. use a zero-percentage threshold) and recompute the residuals which proved to be sufficient in some cases. This blurring strategy is not a necessary requirement for DoB-CLEAN, but an alternative way to guide the imaging similar to how it is done with tapers in CLEAN. But it is translated in the image domain due to the simple possibility to readd any missing small-scale structure at later point in the iterations.

\section{Tests on Synthetic Data}\label{sec: synthetic_test}

\subsection{Synthetic Data}
In the following we test our imaging algorithm on several test images. For these purpose we choose a range of test images presenting various source structures and uv-coverages: we study a synthetic image with a Gaussian core and faint ellipse observed with EVN coverage (\textit{gaussian-evn}), a double-sided core dominated synthetic source with a synthetic ring-like uv-coverage (\textit{dumbbell-ring}), and a synthetic observation of BL Lac with RadioAstron (\textit{bllac-space}). 

\begin{figure*}
    \centering
    \includegraphics[width=\textwidth]{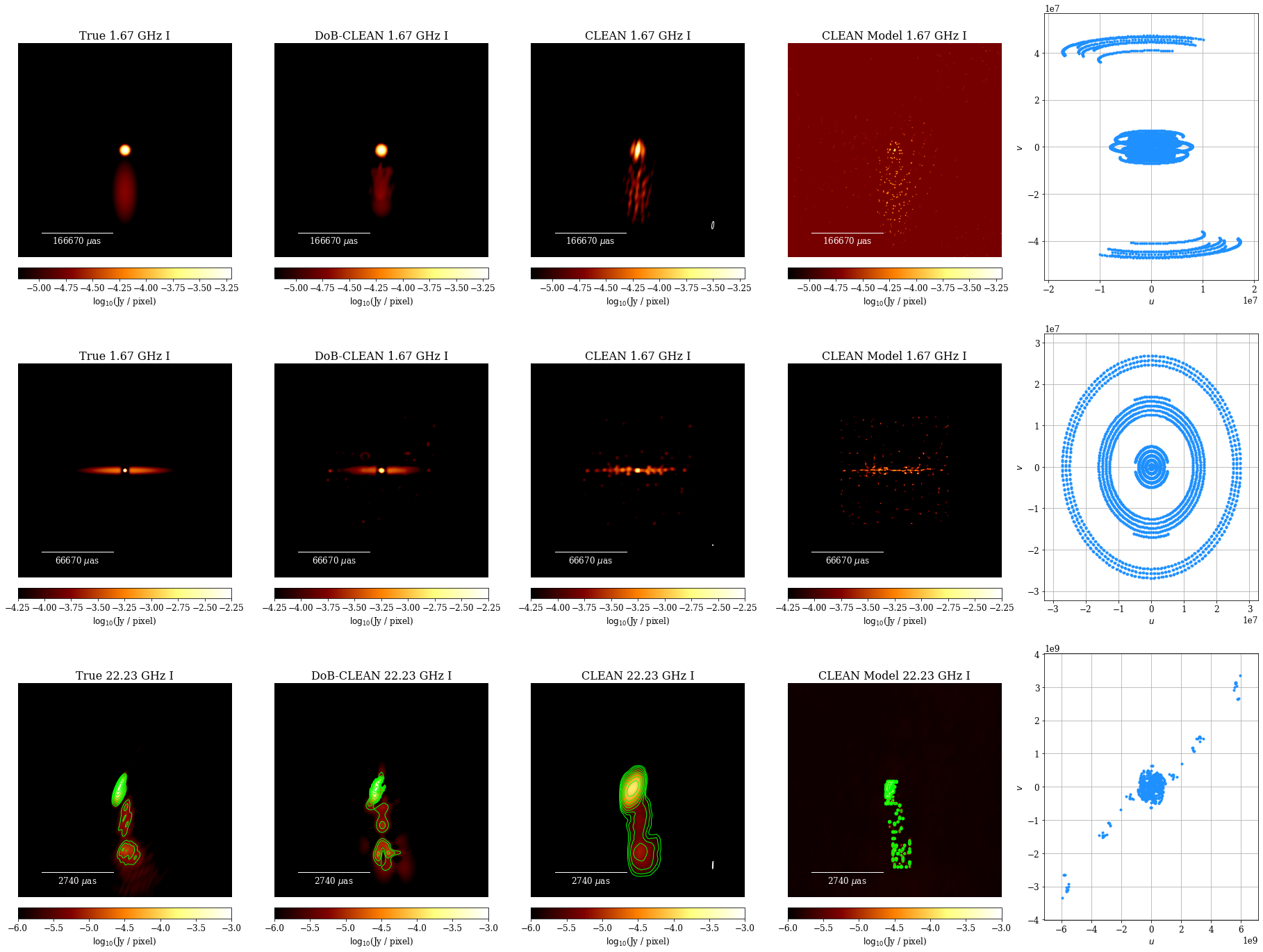}
    \caption{Comparison of reconstructions on synthetic data (first row: \textit{gaussian-evn}, second row: \textit{dumbbell-ring}, third row: \textit{bllac-space}) with various algorithms. First column: true image, second column: DoB-CLEAN reconstruction, third column: CLEAN image, fourth column: CLEAN model, fifth column: uv-coverage of synthetic observation. The contour levels for the \textit{bllac-space} example are $[0.5\%, 1\%, 2\%, 4\%, 8\%, 16\%, 32\%, 64\%]$ of the peak flux.}
    \label{fig: comparison}
\end{figure*}

The \textit{gaussian-evn} model consists of a small Gaussian with width of $5\,\mathrm{mas}$ ($0.5\,\mathrm{Jy}$) and a (faint) elliptical blob with semi-axes of $50\,\mathrm{mas}$ and $20\,\mathrm{mas}$ directed to the south ($0.5\,\mathrm{Jy}$). The elliptical source is shifted by $100\,\mathrm{mas}$ to the south. The \textit{gaussian-evn} model is chosen to artificially approximate typical core-jet structures. The model is plotted in the first panel of Fig. \ref{fig: comparison}. We synthetically observe the model with a past EVN configuration from \citet{Lobanov2011} and observed the synthetic source by the software \textit{ehtim} \citep{Chael2018} with the observe\_same option. The uv-coverage of this observation is plotted in panel five of Fig. \ref{fig: comparison}.

The \textit{dumbbell-ring} model consists of an ellipse with $50\,\mathrm{mas}$ times $500\,\mathrm{mas}$ semi-axes ($1\,\mathrm{Jy}$) centered at the middle, a Gaussian with width $2\,\mathrm{mas}$ ($0.3\,\mathrm{Jy}$) and a second negative Gaussian with with $5\,\mathrm{mas}$ ($-0.3\,\mathrm{Jy}$). The Gaussians and ellipse were chosen in a way that no negative flux appears in the model. The source model is presented in panel 1 of row 2 of Fig. \ref{fig: comparison}. We observed the source for testing purposes with a synthetic instrument with ring-like uv-coverage; for this reason we placed artificial antennas equally spaced from the south pole, observed the synthetic source and flagged out all baselines that did not involve the central station. From this uniform uv-distribution we then introduced two significant radial gaps by flagging. The corresponding uv-coverage is presented in Fig. \ref{fig: comparison}, panel 5 of row 2.

Finally, we took RadioAstron observations of BL Lac as a more physical source model. We took the natural weighted image from \citet{Gomez2016} as the true source structure (see panel 1 row 3 in Fig. \ref{fig: comparison}) and observed it, again with the observe\_same option, with the array of that observation. The corresponding (time-averaged) uv-coverage is plotted in Fig. \ref{fig: comparison}, panel 5 row 3. 

All the observations had thermal noise added, but without adding phase or amplitude calibration errors.

\begin{figure*}
    \centering
    \includegraphics[width=\textwidth]{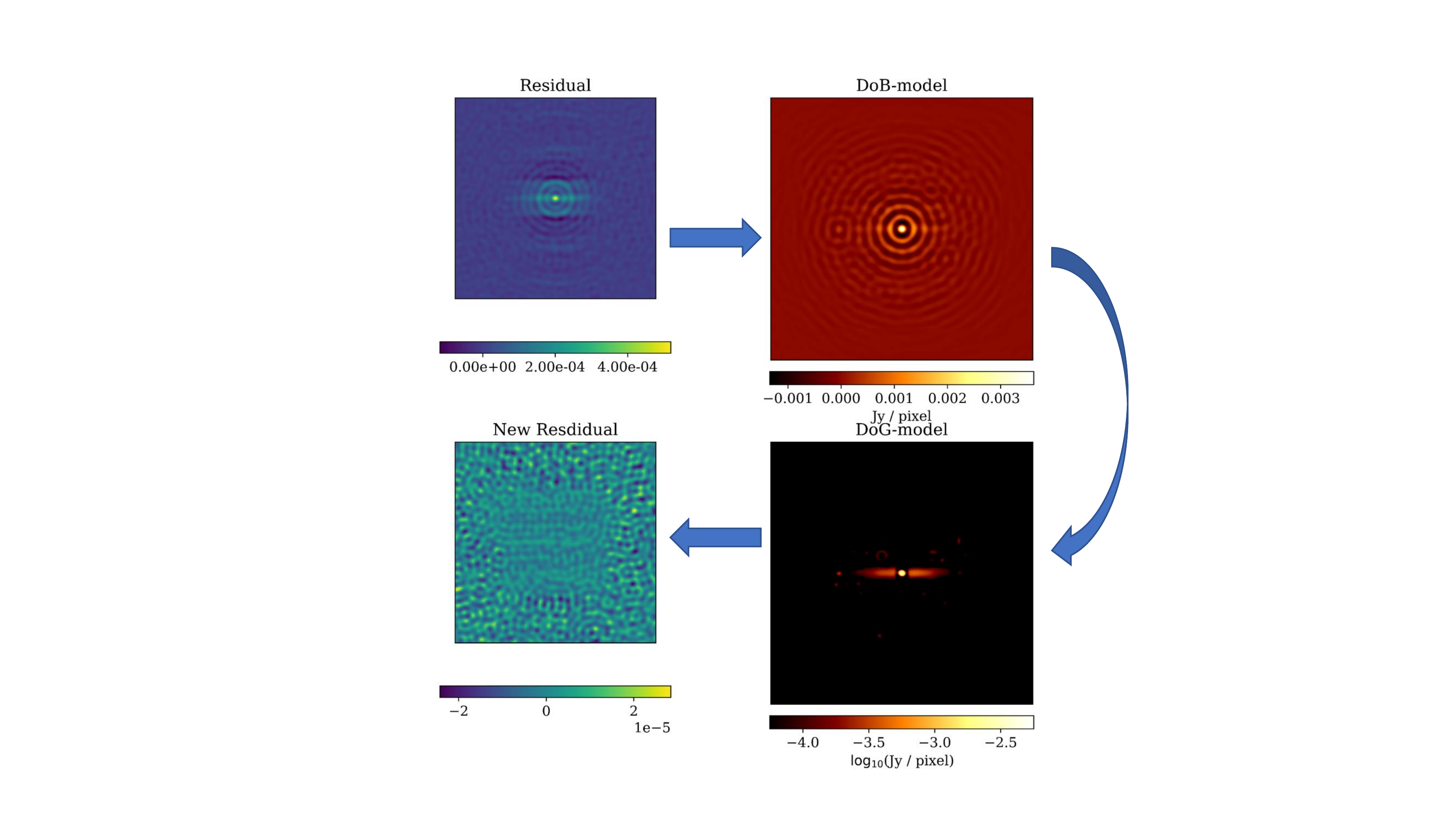}
    \caption{Sketch of the imaging iterations for the \textit{dumbbell-ring} example. Upper Left: Initial residual. Upper right panel: We remove the dirty beam by computing a multiscalar model composed of DoB-wavelets. The panel presents the recovered model $\sum g \cdot I_j^k \cdot \Psi_{clean}^{DoB} \delta_j^k$ (notation from Tab. \ref{alg: wclean}). Bottom right panel: We replace the DoB-wavelets by DoG-wavelets: $\sum g \cdot I_j^k \cdot \Psi_{clean}^{DoB} \delta_j^k$ getting a physically reasonable model that still fits the data. Bottom left: Final updated residual computed from the DoG-model. Iterations continue if needed.}
    \label{fig: sketch_dobclean}
\end{figure*}

\begin{figure*}
    \centering
    \includegraphics[width=\textwidth]{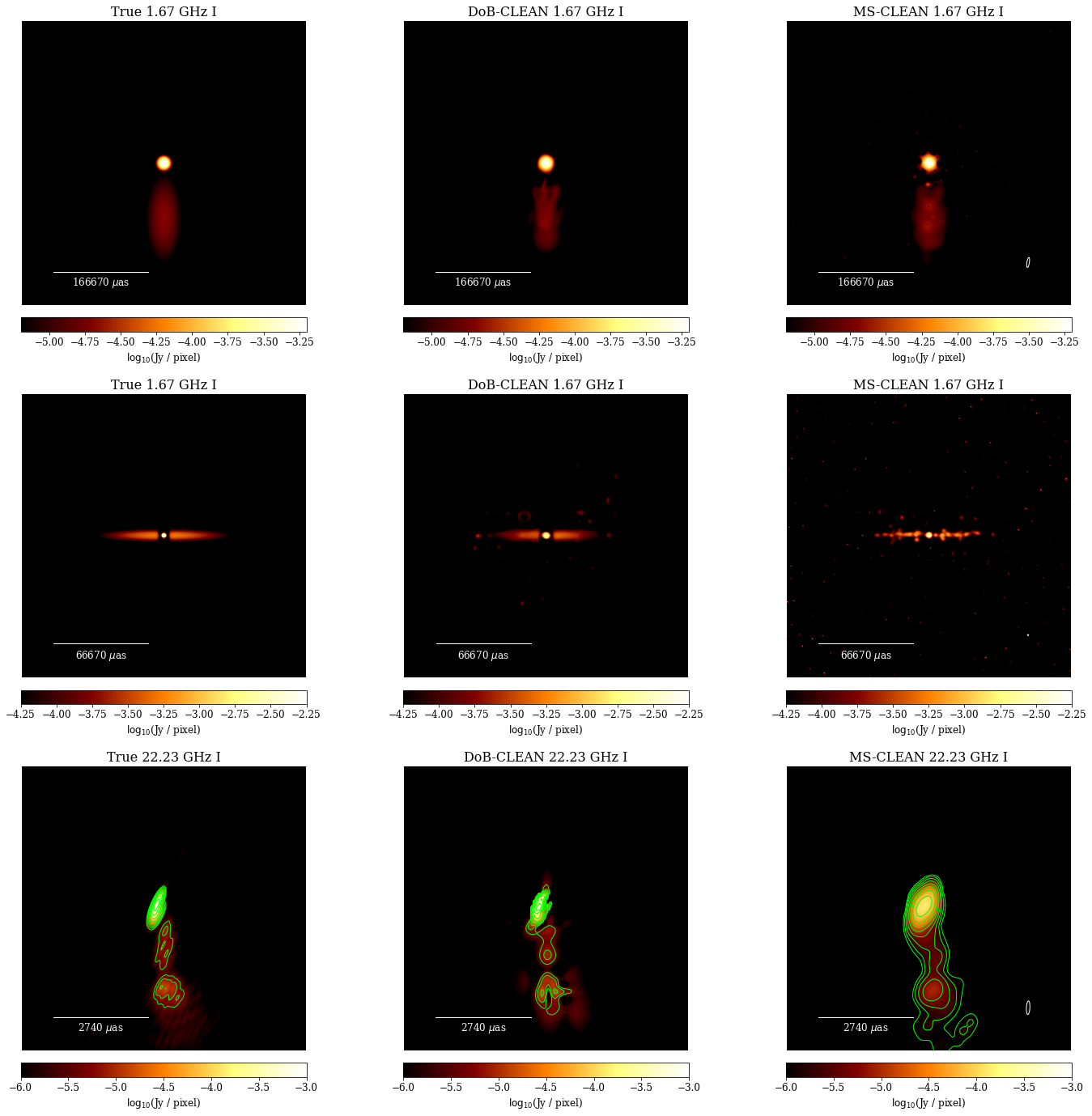}
    \caption{Comparison of reconstructions on synthetic data (first row: \textit{gaussian-evn}, second row: \textit{dumbbell-ring}, third row: \textit{bllac-space}) with various algorithms. First column: true image, second column: DoB-CLEAN reconstruction, third column: MS-CLEAN image. The contour levels for the \textit{bllac-space} example are $[0.5\%, 1\%, 2\%, 4\%, 8\%, 16\%, 32\%, 64\%]$ of the peak flux.}
    \label{fig: comparison_ms}
\end{figure*}

\subsection{Qualitative Comparison} \label{ssec: comparison}
Fig. \ref{fig: comparison} presents the reconstructions of our three synthetic sources with DoB-CLEAN (second column) and with CLEAN (third column: final CLEAN image, fourth column: CLEAN model). For the \textit{bllac-space} model a set of rectangular windows that constrain the flux to the lower half of the image was used. For the \textit{gaussian-evn} and the \textit{dumbbell-ring} reconstructions no particular window was used. Fig. \ref{fig: sketch_dobclean} presents an outline for the imaging procedure done for the \textit{dumbbell-ring} example. We remove the dirty beam successfully during the minor loop, but represent the image by a multiscalar set of DoB-wavelets that contain sidelobes on its own. By replacing the DoB-wavelets by DoG-wavelets we get a physically meaningful from which we recompute a significantly reduced residual.

We show additionally in Fig. \ref{fig: comparison_ms} a comparison of the DoB-CLEAN reconstruction with MS-CLEAN reconstructions. For the MS-CLEAN reconstructions we used in all three examples a dictionary consisting of a delta component and Gaussians with one, two and three times the width of the clean beam.

The DoB-CLEAN reconstructions were very successful overall. The core-jet-like structures were well represented, even if the array configuration was extremely sparse. The representation of the wider, extended emission, in particular in the \textit{gaussian-evn} example is excellent, opposed to CLEAN. As expected a similar effect is achieved by MS-CLEAN reconstructions opposed to H\"{o}gbom CLEAN (compare the upper panels in Fig. \ref{fig: comparison} and \ref{fig: comparison_ms}. The reconstructions of the wide-field \textit{gaussian-evn} structure in Fig. \ref{fig: comparison_ms} is of similar quality between DoB-CLEAN and MS-CLEAN. Moreover, the DoB-CLEAN reconstruction allows for the reconstruction of small scales simultaneously as demonstrated with the two-component core in the \textit{bllac-space} image (indicating a good use of space-baselines).

When comparing to CLEAN (third column) it becomes obvious that DoB-CLEAN achieves super-resolution. It reliably recovers structures smaller than the clean beam, in particular in the \textit{bllac-space} example, even if these structures are faint compared to the central core region (fainter by a factor $\approx 100-1000$ for \textit{bllac-space}). This super-resolving feature, however, does not come at the price of reduced sensitivity to extended emission as discussed above. MS-CLEAN reconstructions are bound to the clean beam resolution as well, hence being outperformed by DoB-CLEAN in terms of resolution as well.

We present in the fourth column of Fig. \ref{fig: comparison} the single CLEAN model, i.e. the composition of delta components. Recall that we identified the mismatch between the final image and the CLEAN model that fits the data as a main theoretical disadvantage of CLEAN. The same applies for MS-CLEAN. In fact, the model maps are no useful description of the source structure in either way. DoB-CLEAN directly computes a model with physical meaning. The reconstructions shown here match the model fitted to the visibilities. Hence, the cleaning with DoB-CLEAN leaves a similar final residual (dominated by thermal noise) as the standard H\"{o}gbom CLEAN, but with a much more useful source model. In this sense, DoB-CLEAN produces more robust source structures.

While CLEAN and MS-CLEAN reconstructions are overall quite successful as well, we identify several qualitative metrics in which DoB-CLEAN clearly outperforms CLEAN and MS-CLEAN. All in all, we conclude from here that DoB-CLEAN seems to be an improvement over CLEAN in terms of resolution (achieving super-resolution), robustness (model matches to final image) and sensitivity to extended emission. The latter advantage becomes obvious in particular for the \textit{gaussian-evn} data set in which the CLEAN beam is much smaller than the extended elliptical source structure, leading to a fractured reconstruction opposed to the smooth extended emission recovered by DoB-CLEAN.

\subsection{Performance Tests}
We now use the \textit{gaussian-evn} example for a set of additional tests to study the features and performance of DoB-CLEAN further.

\begin{figure*}
    \centering
    \includegraphics[width=\textwidth]{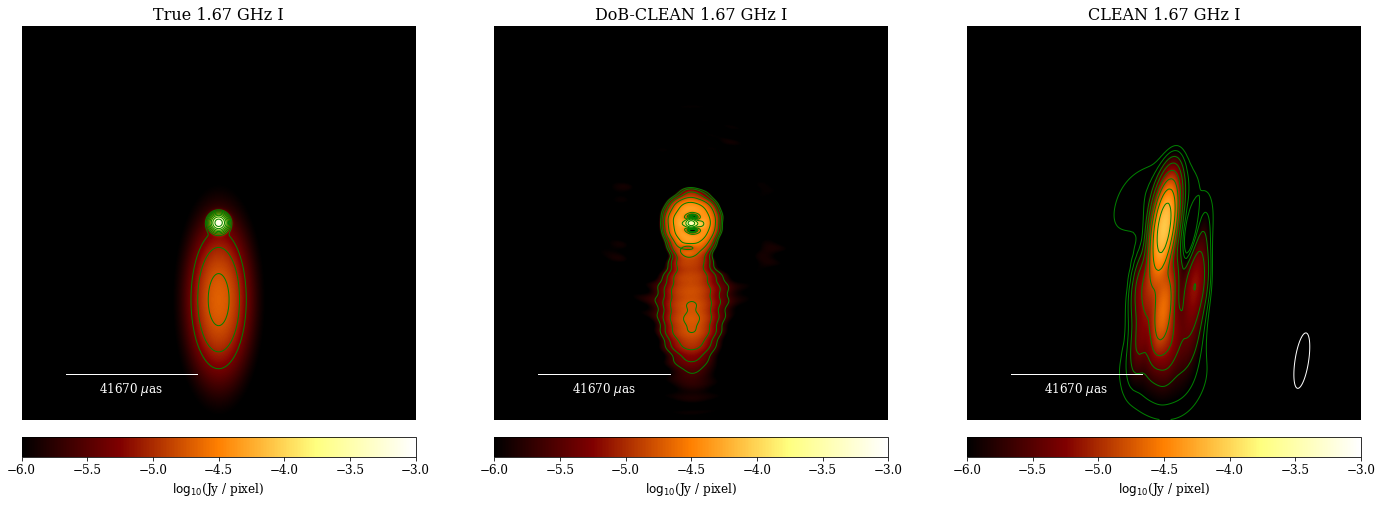}
    \caption{Reconstructions of the \textit{gaussian-evn} test case, but with smaller source size. The contour levels are $[0.5\%, 1\%, 2\%, 4\%, 8\%, 16\%, 32\%, 64\%]$ of the peak flux.}
    \label{fig: gaussian_ellipse_small}
\end{figure*}

To discuss the advantage of super-resolution further, we redid the \textit{gaussian-evn} observation and reconstruction, but with a source structure scaled down by a factor of four in size to highlight the signal on longer baselines more. We present our reconstructions in Fig. \ref{fig: gaussian_ellipse_small}. The extended, elliptical emission is still very well recovered by DoB-CLEAN. The small Gaussian core is overestimated in size due to the large beam size and a smaller central core component becomes visible as a signal from the long baselines. However, the CLEAN reconstruction again has bigger issues with the beam size and the (elliptical) beam shape. This example demonstrates the potential for super-resolving structures at the size of the clean beam with DoB-CLEAN. 

\begin{figure*}
    \centering
    \includegraphics[width=\textwidth]{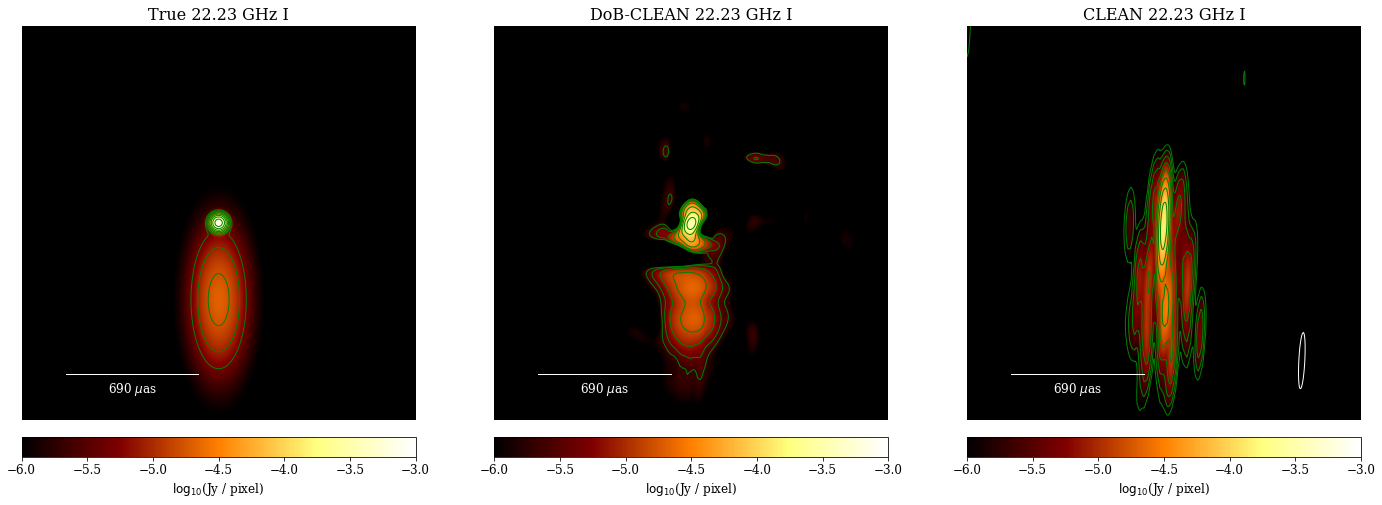}
    \caption{Reconstructions of the \textit{gaussian-evn} test case with RadioAstron uv-coverage. The contour levels are $[0.5\%, 1\%, 2\%, 4\%, 8\%, 16\%, 32\%, 64\%]$ of the peak flux.}
    \label{fig: gaussian_ellipse_radioastron}
\end{figure*}

With this excellent performance at hand for small source structures that require super-resolution, we advance on this statement by studying the \textit{gaussian-evn} source example with synthetic RadioAstron observations (as space-VLBI observations are typically designed to study sources at the highest resolution). VLBI observations with space antennas, however, pose a new range of challenges: the special uv-coverage leading to highly elliptical beams, a bad signal-to-noise ratio on the long space baselines, and the complex calibration of the space baselines. In this study we ignored calibration issues, but we considered highly scale-dependent noise by mirroring a real observation \citep{Gomez2016}. We took the \textit{gaussian-evn} source, scaled it down in size from a field of view of $1''$ to a field of view of $16\,\mathrm{mas}$ (e.g. by a factor of $\approx 16$) and synthetically observed it with RadioAstron. Our reconstructions are shown in Fig. \ref{fig: gaussian_ellipse_radioastron}. This test run again solidifies the problem that CLEAN reconstructions seem to have for highly elliptical beams. DoB-CLEAN works better in this regard, recovers a clearly visible core and a disconnected, approximately elliptical extended emission pattern without many sidelobes. However, compared to the reconstructions that we presented in Fig. \ref{fig: gaussian_ellipse_small} the reconstruction is worse due to the sparsity at small scales (long baselines). The circular Gaussian core-component is represented by a dumb-bell structure instead, the elliptical faint emission is recovered by two connected Gaussian blobs. The dumb-bell structure is a consequence of relative sparsity at small scales as it represents the typical structure that a single scale out of the difference of elliptical beams dictionary features. Basically, only the scale oriented in the direction described by the longer-elongating space baselines is selected, all other scales at this radial width are suppressed. All in all, we can conclude that DoB-CLEAN is capable of reconstructing super-resolved images even with such challenging arrays such as RadioAstron, although a higher level of artifacts is visible at higher resolution.

\begin{figure*}
    \centering
    \includegraphics[width=0.8\textwidth]{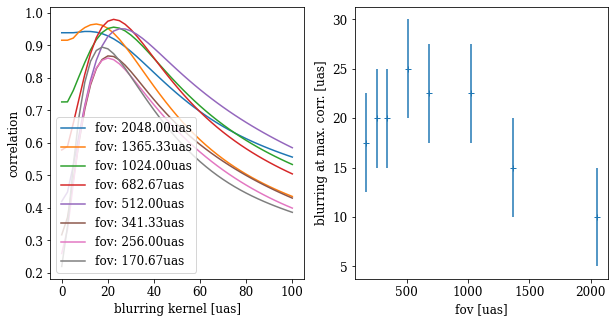}
    \caption{Left panel: Correlation between DoB-CLEAN reconstructions for varying source sizes with RadioAstron synthetic observation of the \textit{gaussian-evn} ground truth image and the blurred ground truth images. Right panel: Blurring at maximal correlation as a function of source size (field of view).}
    \label{fig: resolution_radioastron}
\end{figure*}

\begin{figure*}
    \centering
    \includegraphics[width=0.8\textwidth]{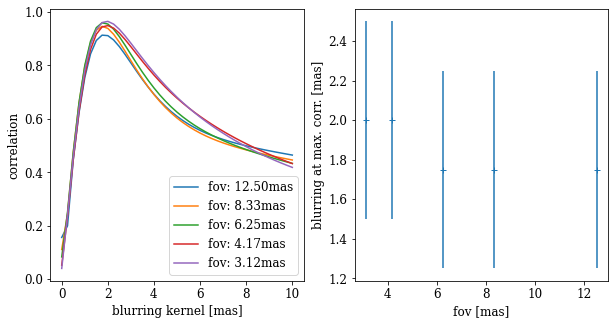}
    \caption{Same as Fig. \ref{fig: resolution_radioastron}, but with EVN coverage.}
    \label{fig: resolution_evn}
\end{figure*}

It is difficult to quantify the amount of super-resolution in general. Since the limiting resolution is not limited by a well-defined beam convolution, but due to the balancing between fitting the visibilities and a multiscale sparsity assumption. The achievable resolution depends both on the specifics of the instrument (i.e. uv-coverage and scale-dependent noise-level) and the source structure itself. To get a rough impression of the resolution that is achievable with DoB-CLEAN we apply the following strategy: we observe the \textit{gaussian-evn} source model with RadioAstron coverage, see Fig. \ref{fig: resolution_radioastron}, and with EVN coverage, see Fig. \ref{fig: resolution_evn}. Iteratively, we minimize the source size (by keeping the same image array, but minimizing the pixel size, i.e. the field of view). Each time we do a reconstruction with DoB-CLEAN and blur the (minimized) ground truth images on a predefined fine grid of circular Gaussian blurring kernels. We compute the correlation of the blurred synthetic ground truth images and the reconstructions in any case (left panels in Fig. \ref{fig: resolution_radioastron} and Fig. \ref{fig: resolution_evn}). The correlation curves look reasonable with a clearly identifiable maximal peak. We show the blurring kernel size with the maximal correlation for the smallest source sizes in the right panels. If the source is that small that it becomes unresolved by DoB-CLEAN, the blurring kernel size needs to converge from below roughly towards the limiting resolution: indeed the maximum correlation is roughly constant within the errorbars indicating an effective resolution for a RadioAstron configuration of $\sim 20\,\mu\mathrm{as}$ (beam: $\sim 290 \times 31\,\mu\mathrm{as}$) and an effective resolution for an EVN resolution of $\sim 2\,\mathrm{mas}$ (beam: $\sim 18\times 4\,\mathrm{mas}$). Hence, moderate super-resolution by a factor of $2-3$ might be possible. However, while the representation of super-resolved features with wavelets is clearly more reasonable than a representation with delta components, we have to note that the reconstruction problem at a higher resolution is also more challenging and artifacts that are usually hidden under the convolution with the beam can be expected (and are visible for example in Fig. \ref{fig: gaussian_ellipse_radioastron}).

\begin{figure*}
    \centering
    \includegraphics[width=\textwidth]{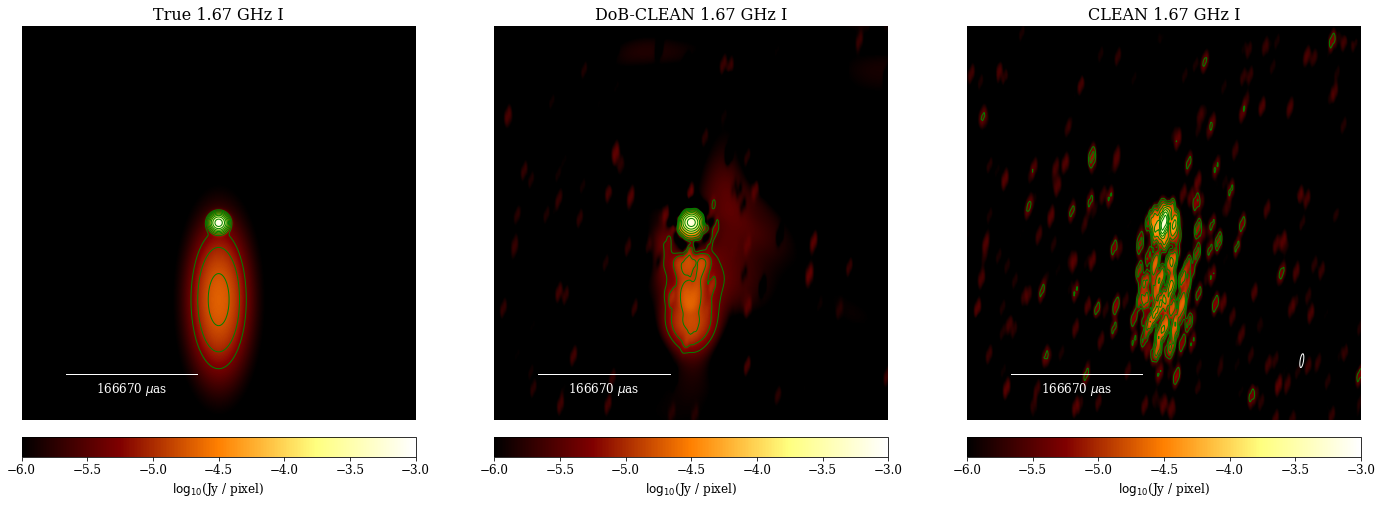}
    \caption{High dynamic range reconstructions of the \textit{gaussian-evn} test case, but higher thermal noise level. The contour levels are $[0.5\%, 1\%, 2\%, 4\%, 8\%, 16\%, 32\%, 64\%]$ of the peak flux.}
    \label{fig: gaussian_ellipse_high_noise}
\end{figure*}

Finally, we study the effect of thermal noise on the reconstruction. For this purpose we again observed the \textit{gaussian-evn} example, but this time added a constant thermal noise on all baselines at a level such that the final signal-to-noise ratio is approximately one. The reconstructions are presented in Fig. \ref{fig: gaussian_ellipse_high_noise}. Comparing the reconstruction shown in Fig. \ref{fig: comparison}, the source structures recovered by DoB-CLEAN and CLEAN remain relatively unaffected. Faint, blobby background sidelobes as expected from Gaussian noise are introduced to the CLEAN image. In DoB-CLEAN the effect is different: a coronal emission around the central component is introduced. This feature, however, is very weak and can only be seen at high dynamic range. This coronal feature has to be noted as an explicit image artifact that DoB-CLEAN introduces in the image when studying noisy images at high dynamic range. 

\subsection{Artifacts compared to CLEAN} \label{ssec: artifacts}

\begin{figure*}
    \centering
    \includegraphics[width=\textwidth]{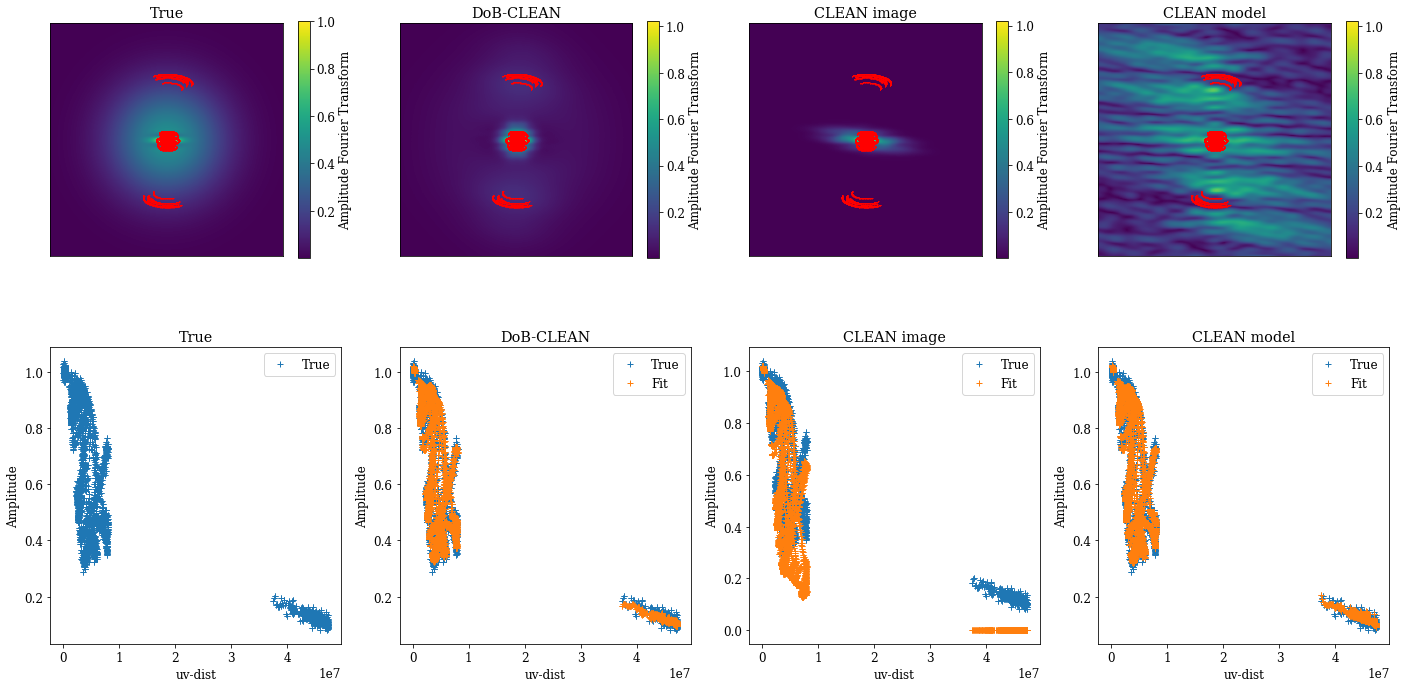}
    \caption{Comparison of the fits of the \textit{gaussian-evn} synthetic data reconstruction in the Fourier domain. Upper panels: complete Fourier transform of reconstructions (true, DoB-CLEAN, CLEAN image and CLEAN model) with uv-coverage over-plotted (red crosses). Lower panels: Radplot showing the fit of the recovered model to the observed visibilities. Only for DoB-CLEAN the fit is successful (lower panels) and the Fourier transform of the model is physically reasonable (upper panels).}
    \label{fig: visibility_fit}
\end{figure*}

We now compare DoB-CLEAN to CLEAN with the \textit{gaussian-evn} example with a reduced source size, see Fig. \ref{fig: gaussian_ellipse_small} with special emphasis on the image artifacts introduced by these algorithms. We present the complete Fourier transform of the true image and the reconstructions (DoB-CLEAN, CLEAN model and final CLEAN image) in Fig. \ref{fig: visibility_fit}. In the upper row we show the amplitude of the Fourier transform of the true source model and the uv-points over-plotted in red. In the lower panels we show the fit between the measured and observed visibilities. Standard imaging software such as \textit{Difmap} typically only show the latter ones indicating a successful fit of the observed visibilities for both CLEAN (i.e. the model) and DoB-CLEAN. However, the complete Fourier transform reveals that this might be inadequate. The CLEAN reconstruction shows a rich, periodic structure in the Fourier domain, in the gap between short and long baselines, but also at baselines longer than the observed ones. These structures in the gaps are not motivated by any measured visibility and in particular correlate very little with the signal measured at long baselines. This particular CLEAN problem is solved by convolving with the clean beam, but at the cost of a worsened fit of the final image to the observed visibilities, compare the bottom panel for the CLEAN image. The DoB-CLEAN reconstruction shows a much better fit to the Fourier coefficients. The signal in the large gap between short and long baselines is suppressed as also is the unphysical signal on Fourier coefficients longer than the longest baselines, but the fit to the observed baselines remains excellent.

\begin{figure*}
    \centering
    \includegraphics[width=\textwidth]{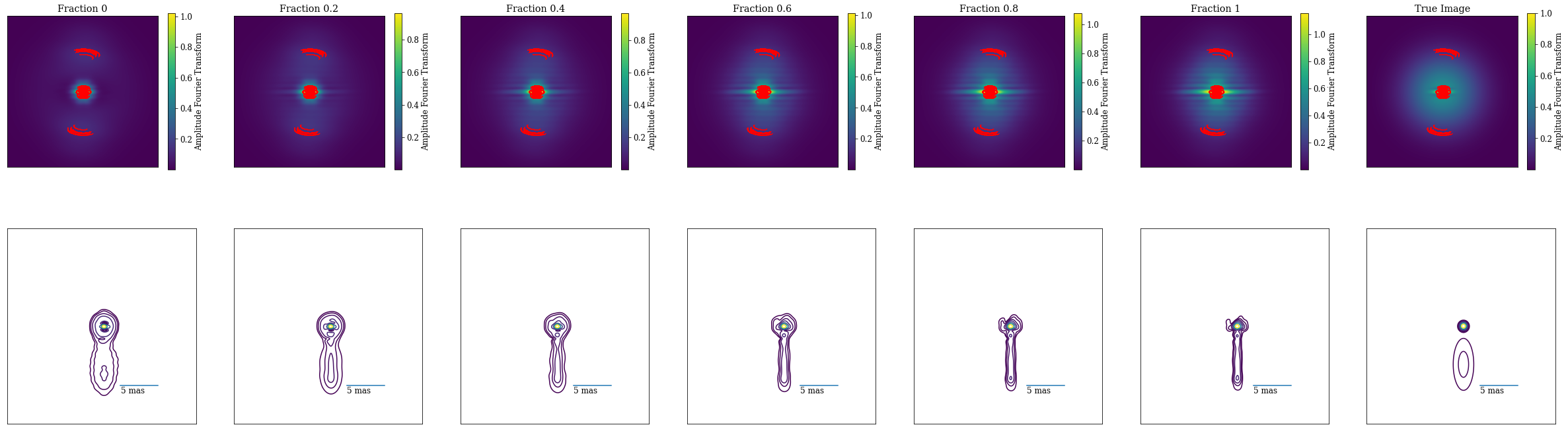}
    \caption{Fourier transform of recovered data with DoB-CLEAN (upper panels) and the recovered model (lower panels) in the \textit{gaussian-evn} test case. From left to right the missing (not measured) scales are interpolated from the covered scales with a higher fraction. The most right panels show the true image. The used contour-levels are $[1\%, 2\%, 4\%, 8\%, 16\%, 32\%, 64\%]$ of the peak flux.}
    \label{fig: add_missing_scales}
\end{figure*}

Due to this suppression, minimal structural information is added in the gaps and only the robust, measured image information is processed. However, comparing to the true Fourier transform, this also gives rise to some possible problems in the imaging procedure: as the uv-coverage is sparse and contains a prominent gap with unmeasured Fourier coefficients, there is image information in this gap that is not recovered in the final image with DoB-CLEAN. In particular this gap introduces the spurious image structure visible in Fig. \ref{fig: gaussian_ellipse_small} in the core component. The core Gaussian is recovered by a small DoG component compressed by the longest baselines in the array, and a wider Gaussian component compressed by the shorter baselines. The missing scale (i.e. a missing DoG-scale to satisfy completeness) is visible in the final image by the ring-like feature of weak flux sources around the inner component. While imaging only robust image features with a reduced sidelobe level sounds like an optimal solution for imaging, these kinds of structures are a clear indicator of missing amplitudes on non-measured baselines. As explained in Sec. \ref{ssec: postprocessing} DoB-CLEAN, as opposed to CLEAN, offers a unique way to identify these problems and to re-add these uncovered scales in the image. We demonstrate the usefulness of this approach in Fig. \ref{fig: add_missing_scales}. With an increasing fraction of added missing scales, the interpolated flux in the gap becomes more prominent (upper panels). The artifact in the core component vanishes (bottom panels). When overdoing the interpolation however (i.e. adding too much information on small scales/long baselines), the elliptic extended emission gets wrongly estimated. Hence, on observational data this interpolation option should be used with relative caution as we are interpolating structural information in the image that is in principle unmeasured. 

\section{BL Lac} \label{sec: bllac}

\subsection{Data}
We reanalyzed the public data set of BL Lac observations with RadioAstron \citep{Gomez2016} in this section as an additional test with real observational data. In what follows we summarize these observations, for more detailed information we refer to \citet{Gomez2016}. BL Lac was observed at $22\,\mathrm{GHz}$ on 10 November and 11 November 2013. Due to some technical problems BL Lac was only observed by 15 correlated antennas (instead of the 26 possible in the array). The data set was correlated at the DiFX correlator at the Max-Planck-Institut für Radioastronomie (MPIfR). Data reduction and calibration took place with AIPS and Difmap \citep{Shepherd1997}. We used the self-calibrated data set of \citet{Gomez2016} as a starting point for reconstructions with DoB-CLEAN.

\subsection{Reconstructions}
\begin{figure*}
    \centering
    \begin{subfigure}{}
         \centering
         \includegraphics[width=0.7\textwidth]{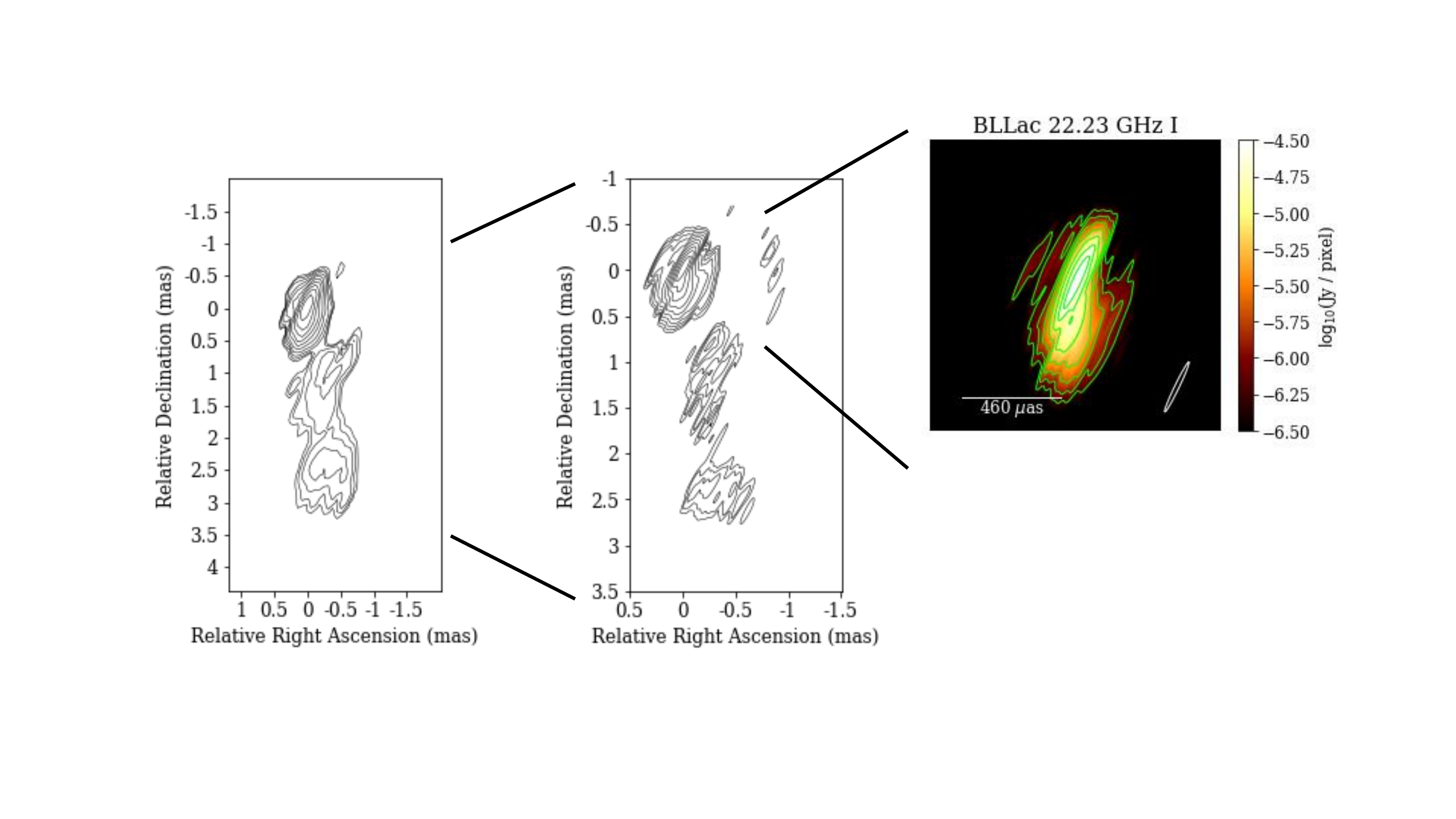}
         \caption{BL Lac reconstructions with DoB-CLEAN blurred with the clean beam with natural weighting (left panel), uniform weighting (middle panel) and uniform weighting with smaller pixel size zoomed in the central core region (right panel). The contour levels are: [0.1,0.2,0.4,0.8,1.6,3.2,6.4,12.8,25.6,51.2]\% ([0.1,0.2,0.4,0.8,1.6,3.2,6.4,12.8,25.6,51.2]\%, [0.8,1.6,3.2,6.4,12.8,25.6,51.2]\%) of the respective peak brightness.}
         \label{fig: clean}
    \end{subfigure}
    \begin{subfigure}{}
         \centering
         \includegraphics[width=0.7\textwidth]{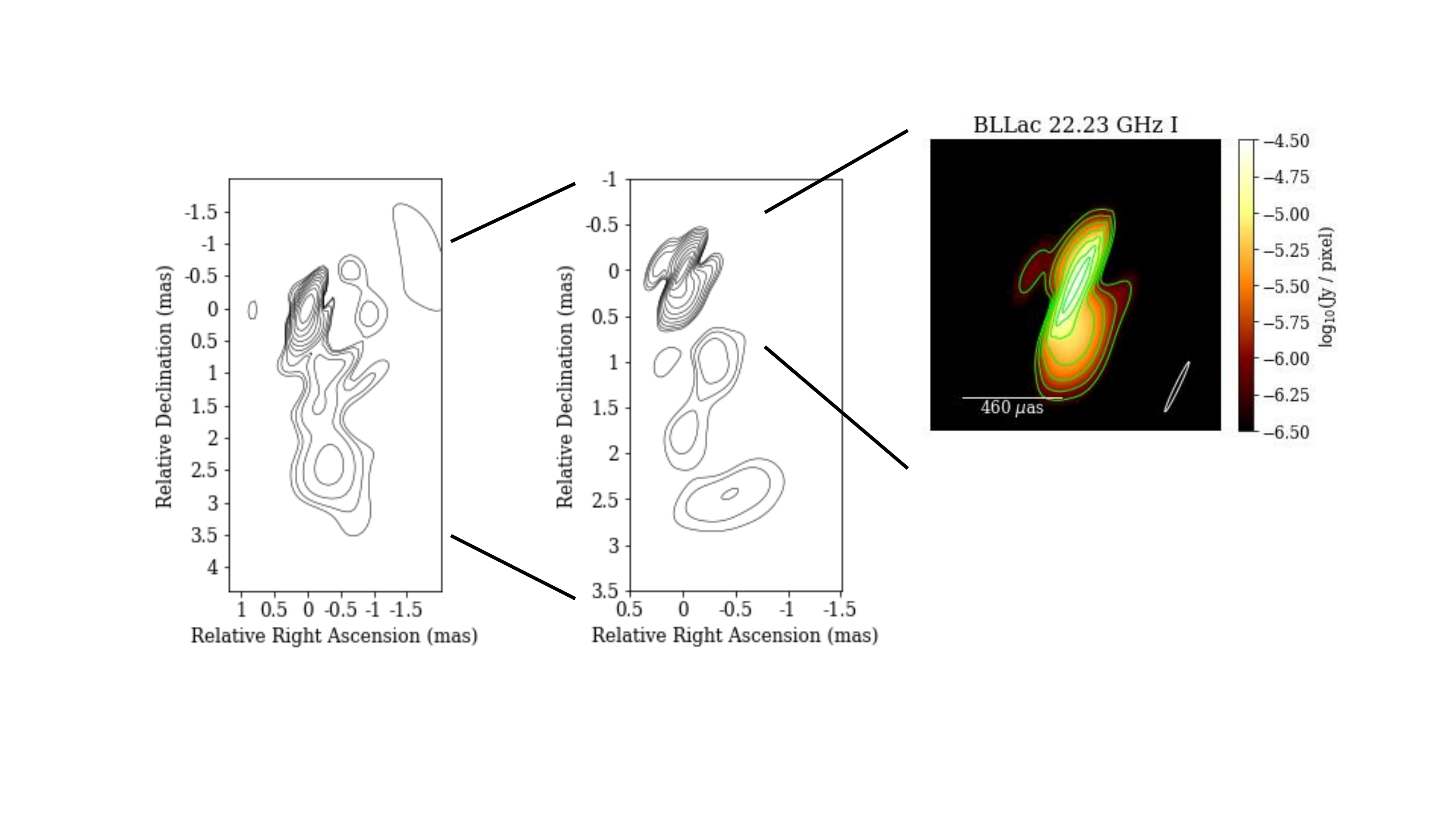}
         \caption{BL Lac reconstructions with DoB-CLEAN blurred with the clean beam with natural weighting (left panel), uniform weighting (middle panel) and uniform weighting with smaller pixel size zoomed in the central core region (right panel). The contour levels are: [0.1,0.2,0.4,0.8,1.6,3.2,6.4,12.8,25.6,51.2]\% ([0.025, 0.05, 0.1,0.2,0.4,0.8,1.6,3.2,6.4,12.8,25.6,51.2]\%, [0.8,1.6,3.2,6.4,12.8,25.6,51.2]\%) of the respective peak brightness.}
         \label{fig: psf}
    \end{subfigure}
    \begin{subfigure}{}
         \centering
         \includegraphics[width=0.7\textwidth]{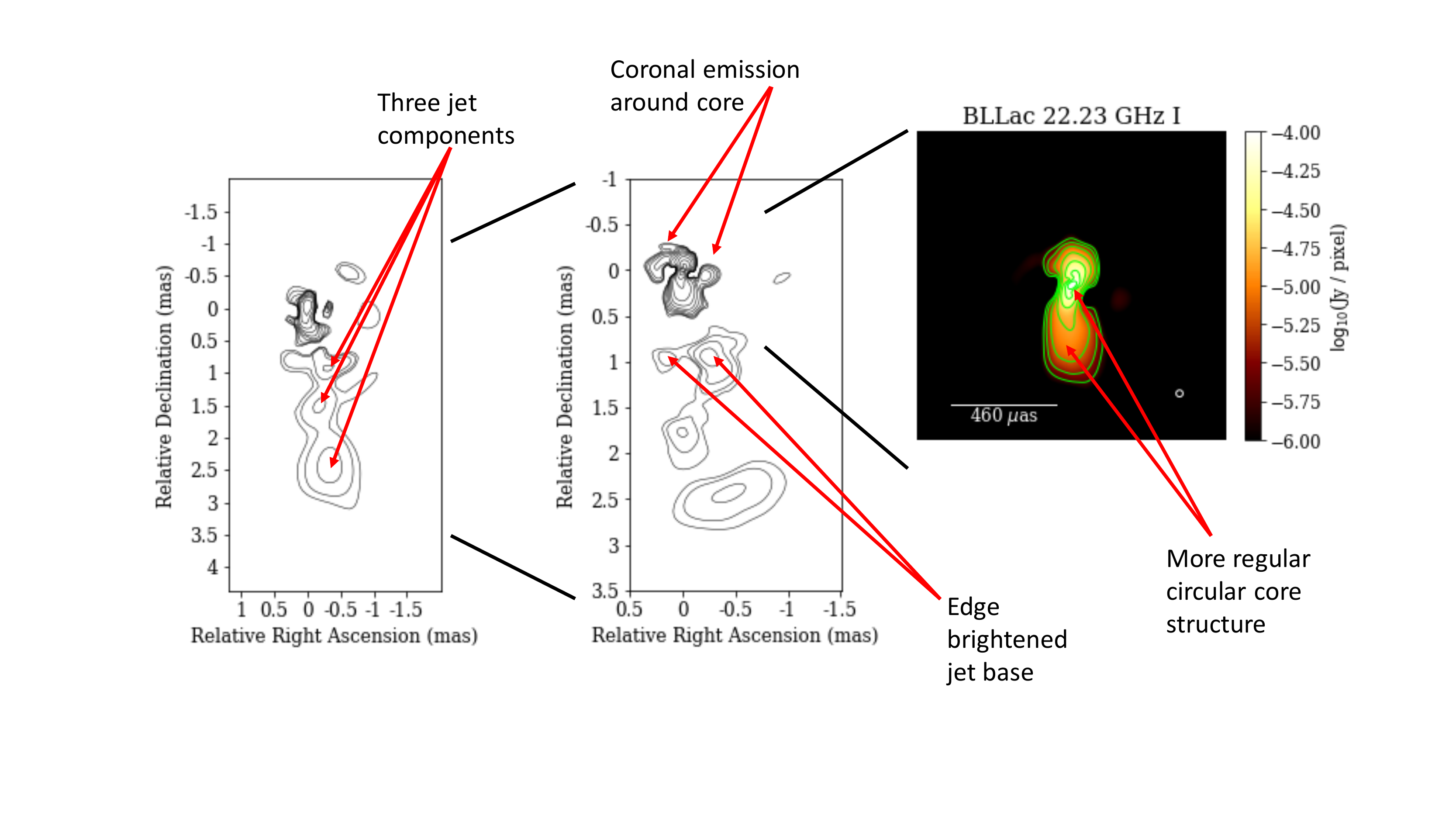}
         \caption{BL Lac reconstructions with DoB-CLEAN at full resolution with natural weighting (left pane), uniform weighting (middle panel) and uniform weighting with smaller pixel size zoomed in the central core region (right panel). The contour levels are: [0.1,0.2,0.4,0.8,1.6,3.2,6.4,12.8,25.6,51.2]\% ([0.025, 0.05, 0.1,0.2,0.4,0.8,1.6,3.2,6.4,12.8,25.6,51.2]\%, [0.8,1.6,3.2,6.4,12.8,25.6,51.2]\%) of the respective peak brightness.}
         \label{fig: full_res}
    \end{subfigure}
\end{figure*}

We present our reconstruction results with DoB-CLEAN in Fig. \ref{fig: full_res}. Moreover, we show our reconstructions blurred with the corresponding clean beam in Fig. \ref{fig: psf}.

Comparing our imaging results blurred with the clean beam (Fig. \ref{fig: psf}) to the reconstruction results with CLEAN (Fig. \ref{fig: clean}), we identify very similar structures, in particular for natural weighting. We identify the central core with an elliptic shape, and the two connected Gaussian blobs to the south. Some of the fine-structure in the CLEAN image is visible in the DoB-CLEAN image as well, such as the shape of the core or the orientation of the components in the jet. However, there are also some slight differences such as the faint emission to the north-east that is not related to the jet. This emission could be an artifact of DoB-CLEAN reconstructions, compare the typical image artifacts that we discussed in Sec. \ref{ssec: comparison} caused by the intrinsic sidelobes in the basis functions. In the middle panels we show the reconstructions with uniform weighting, and in the right panels a zoom-in on the central core region with uniform weighting. These reconstructions with their more highly resolved structures highlight the core region more. Overall the similarity between the blurred DoB-CLEAN images (Fig. \ref{fig: psf}) and the CLEAN images (Fig. \ref{fig: clean}) is great for uniform weighting, in particular in the zoom-in panels into the core. Interestingly, CLEAN finds stronger extended emission. Moreover, we find a possible edge-brightened structure in the reconstructions with DoB-CLEAN that is not apparent in the CLEAN images.

We demonstrated that DoB-CLEAN allows for super-resolution and the actual model computed has a physical model in contrast to CLEAN. We present in Fig. \ref{fig: full_res} the DoB-CLEAN reconstructions at full resolution. In fact, Fig. \ref{fig: full_res} shows more highly resolved structures of a narrow jet. We like to mention some features that become visible in the full resolution DoB-CLEAN reconstructions as opposed to the blurred reconstructions. 

\begin{itemize}
    \item As visible in the natural-weighted image, we can identify three (instead of two) peaks in the jet emission, the central jet component is now resolved.
    \item We observe a core structure of a very narrow central core component surrounded by a wider coronal emission. This structure cannot be seen with CLEAN or DoB-CLEAN at lower resolution as the feature is blurred out by the clean beam. We note that when comparing the reconstructions of the innermost core region, e.g. the right panels in Fig. \ref{fig: clean} and Fig. \ref{fig: psf}, also the CLEAN reconstructions shows signs of a quasi-coronal emission around the core, i.e. emission to the north-west and to the south-east of the central core component. However, comparing to our discussions in Sec. \ref{sec: synthetic_test} it is also possible that this feature is caused by missing scales in the reconstruction. A further analysis of this feature with alternative super-resolving algorithms, i.e. RML algorithms \citep{Chael2018, Mueller2022}, is well desired but left for subsequent works.
    \item We observe a sign of possible edge-brightening in the jet base due to a second component towards the left. This was not observed with CLEAN reconstructions. This structural feature is also visible in the blurred DoB-CLEAN reconstructions, see the middle panel of Fig. \ref{fig: psf}.
    \item The core structure in CLEAN and blurred DoB-CLEAN has a double-elliptic shape, compare the right panels in Fig. \ref{fig: clean} and Fig. \ref{fig: psf}. With the full-resolution DoB-CLEAN reconstructions, we see a more regular, circular reconstruction of the core, with a clearly visible jet basis in the innermost region.
\end{itemize}

While concordance between all reconstructions is overall very high, the novel DoB-CLEAN reconstructions demonstrate some possible features that are different from CLEAN reconstructions, especially at the highest angular resolution. Some of them could be connected to imaging artifacts either by DoB-CLEAN or standard H\"{o}gbom CLEAN. We discuss the robustness of these features in Appendix \ref{app: reliability} in some more detail. In a nutshell, both the possible edge-brightening and the coronal emission around the core could be associated with a common sidelobe pattern. The information which emission is real and which emission is thought of to be caused by sidelobes is highly uncertain. This example highlights once more the need for more variety in the choice of reconstruction methods in VLBI. More work on the innermost jet in BL Lac with more modern Bayesian and RML based methods establishing concordance between various methods is left for subsequent works. 

\section{Conclusion}
We developed the novel multi-scalar imaging algorithm DoB-CLEAN. DoB-CLEAN is based on the framework of CLEAN to still allow the straightforward manual manipulation and calibration of data that has proven successful in the VLBI community for the last decades. However, DoB-CLEAN addresses some pathologies of the CLEAN algorithm: CLEAN has spurious regularization properties, is inadequate to describe extended emission, and introduces a separation between the model fitted to the observed visibilites and the final astronomical image. These pathologies are mainly caused by CLEAN approaching the image as a set of delta functions. DoB-CLEAN basically replaces these CLEAN components by wavelets that compress radial and directional information in parallel. The wavelet dictionary is fitted to the uv-coverage which provides a more data-driven approach to imaging. Sidelobes are suppressed by switching between a wavelet dictionary of steep, orthogonal masks in the Fourier domain and a sidelobe free representation in the image domain. 

We implemented DoB-CLEAN and benchmarked its performance against CLEAN reconstructions on synthetic data. DoB-CLEAN succeeds over CLEAN in terms of resolution and accuracy. It removes the separation between model and image, i.e. DoB-CLEAN fits a model to the uv-coverage that in fact has a physical meaning. The perhaps biggest advantage of DoB-CLEAN however is the control over the fit in the gaps of the uv-coverage offered by the multi-scalar wavelet dictionary. Firstly, this helps to prevent overfitting and fosters image robustness (i.e. only measured, robust image features are measured). Secondly, this offers rich opportunities for post-processing, i.e. identifying missing scales and missing image features in the observation and imaging. These post-processing capabilities are also of general interest as they offer a way to identify an uncertainty estimate of cleaned features in VLBI observations. 

Despite these great advantages, DoB-CLEAN does not solve any problem related to sparsity of the uv-coverage. The lack of certain scales in the observation  can introduce artifacts in the DoB-CLEAN imaging results when completely suppressed. Moreover, the basis functions have negative flux that is, on a low level, still present in the final images (i.e. the dynamic range remains limited).

Finally, we applied DoB-CLEAN to some old, already calibrated data from RadioAstron observations of BL Lac. The reconstructions with DoB-CLEAN and with CLEAN share a lot of similarity when blurred to the same resolution, but there are also some differences visible that may alter the scientific interpretation, especially at the highest resolution. This, once more, elucidates the need for more variety in the imaging algorithms used in frontline VLBI observations to establish concordance between them and robustness of the scientific interpretation. We will address this issue in subsequent works. 

\bibliography{lib}{}
\bibliographystyle{aa}

\clearpage
\appendix

\section{Dictionaries} \label{app: dictionaries}

The dictionaries used in this paper are as follows::
\begin{align*}
    \Psi^{DoG}: I \mapsto [&G^r_{\sigma_0} * I - G^e_{\sigma_0, \sigma_{1}, \alpha_0} * I, G^r_{\sigma_0} * I - G^e_{\sigma_0, \sigma_{1}, \alpha_1} * I, ..., G^r_{\sigma_0} * I - G^e_{\sigma_0, \sigma_{1}, \alpha_{N-1}} * I, \frac{1}{B_0} \sum_{i=0}^{N-1}  G^e_{\sigma_0, \sigma_{1}, \alpha_{i}} * I - G^r_{\sigma_1} * I, \\
    &G^r_{\sigma_1} * I - G^e_{\sigma_1, \sigma_{2}, \alpha_{0}} * I, \hspace{1.6cm} ... \hspace{1.6cm}, G^r_{\sigma_1} * I - G^e_{\sigma_1, \sigma_{2}, \alpha_{N-1}} * I, \frac{1}{B_1} \sum_{i=0}^{N-1}  G^e_{\sigma_1, \sigma_{2}, \alpha_{i}} * I - G^r_{\sigma_2} * I,\\
    &G^r_{\sigma_2} * I - G^e_{\sigma_2, \sigma_{3}, \alpha_{0}} * I, \hspace{1.6cm} ... \hspace{1.6cm}, G^r_{\sigma_2} * I - G^e_{\sigma_2, \sigma_{3}, \alpha_{N-1}} * I, \frac{1}{B_2} \sum_{i=0}^{N-1}  G^e_{\sigma_2, \sigma_{3}, \alpha_{i}} * I - G^r_{\sigma_3} * I,\\
    & \vdots \\
    &G^r_{\sigma_{J-1}} * I - G^e_{\sigma_{J-1}, \sigma_{J}, \alpha_{0}} * I, \hspace{1.6cm} ... \hspace{1.6cm}, G^r_{\sigma_{J-1}} * I - G^e_{\sigma_{J-1}, \sigma_{J}, \alpha_{N-1}} * I, \frac{1}{B_{J-1}} \sum_{i=0}^{N-1}  G^e_{\sigma_{J-1}, \sigma_{J}, \alpha_{i}} * I - G^r_{\sigma_J} * I,\\
    &G^r_{\sigma_J} * I]
\end{align*}

and:

\begin{align*}
    \Psi^{DoB}: I \mapsto [&\tilde{J}^r_{\tilde{\sigma}_0} * I - G^e_{\tilde{\sigma}_0, \tilde{\sigma}_{1}, \alpha_0} * I, \tilde{J}^r_{\tilde{\sigma}_0} * I - \tilde{J}^e_{\tilde{\sigma}_0, \tilde{\sigma}_{1}, \alpha_1} * I, ..., \tilde{J}^r_{\tilde{\sigma}_0} * I - \tilde{J}^e_{\tilde{\sigma}_0, \tilde{\sigma}_{1}, \alpha_{N-1}} * I, \frac{1}{B_{0}} \sum_{i=0}^{N-1}  \tilde{J}^e_{\tilde{\sigma}_0, \tilde{\sigma}_{1}, \alpha_{i}} * I - \tilde{J}^r_{\tilde{\sigma}_1} * I, \\
    &\tilde{J}^r_{\tilde{\sigma}_1} * I - \tilde{J}^e_{\tilde{\sigma}_1, \tilde{\sigma}_{2}, \alpha_{0}} * I, \hspace{1.6cm} ... \hspace{1.6cm}, \tilde{J}^r_{\tilde{\sigma}_1} * I - \tilde{J}^e_{\tilde{\sigma}_1, \tilde{\sigma}_{2}, \alpha_{N-1}} * I, \frac{1}{B_1} \sum_{i=0}^{N-1}  \tilde{J}^e_{\tilde{\sigma}_1, \tilde{\sigma}_{2}, \alpha_{i}} * I - \tilde{J}^r_{\tilde{\sigma}_2} * I,\\
    &\tilde{J}^r_{\tilde{\sigma}_2} * I - \tilde{J}^e_{\tilde{\sigma}_2, \tilde{\sigma}_{3}, \alpha_{0}} * I, \hspace{1.6cm} ... \hspace{1.6cm}, \tilde{J}^r_{\tilde{\sigma}_2} * I - \tilde{J}^e_{\tilde{\sigma}_2, \tilde{\sigma}_{3}, \alpha_{N-1}} * I, \frac{1}{B_2} \sum_{i=0}^{N-1}  \tilde{J}^e_{\tilde{\sigma}_2, \tilde{\sigma}_{3}, \alpha_{i}} * I - \tilde{J}^r_{\tilde{\sigma}_3} * I,\\
    & \vdots \\
    &\tilde{J}^r_{\tilde{\sigma}_{J-1}} * I - \tilde{J}^e_{\tilde{\sigma}_{J-1}, \tilde{\sigma}_{J}, \alpha_{0}} * I, \hspace{1.6cm} ... \hspace{1.6cm}, \tilde{J}^r_{\tilde{\sigma}_{J-1}} * I - \tilde{J}^e_{\tilde{\sigma}_{J-1}, \tilde{\sigma}_{J}, \alpha_{N-1}} * I, \frac{1}{B_{J-1}} \sum_{i=0}^{N-1}  \tilde{J}^e_{\tilde{\sigma}_{J-1}, \tilde{\sigma}_{J}, \alpha_{i}} * I - \tilde{J}^r_{\tilde{\sigma}_J} * I,\\
    &\tilde{J}^r_{\tilde{\sigma}_J} * I]
\end{align*}
where $G^r_\sigma$ denotes a two-dimensional radially symmetric (normalized) Gaussian function with standard deviation $\sigma$ and $G^e_{\sigma_1, \sigma_2, \alpha}$ a two-dimensional elliptical (and normalized) Gaussian function with minor axis $\sigma_1$ and major axis $\sigma_2$ that is rotated by an angle $\alpha$. $\tilde{J}^r_\sigma$ is a two-dimensional radially symmetric modified Bessel function, i.e. $\tilde{J}^r_\sigma (r) = \frac{1}{\sigma r} J_1(2 \pi r / \sigma)$ and $\tilde{J}^e_{\sigma_1, \sigma2, \alpha}$ is the elliptical analog with minor axis minor axis $\sigma_1$ and major axis $\sigma_2$ and rotation angle $\alpha$.

\clearpage
\section{Proof for Selection Criterion} \label{app: proof}

It is:
\small
\begin{align} \nonumber
&\underset{i,m,k}{\mathrm{argmax}} \frac{1}{\norm{\Phi_{i,m} * B^D} \norm{\Phi_{i,m}} } \left( \Phi_{i,m} * B^D * I^D \right) (k) \\ \nonumber
&= \underset{i,m,k}{\mathrm{argmax}} \frac{1}{\norm{\Phi_{i,m} * B^D} \norm{\Phi_{i,m}} } \left( \Phi_{i,m} * B^D * B^D * I \right) (k) \\ \nonumber
&= \underset{i,m,k}{\mathrm{argmax}} \frac{1}{\norm{\Phi_{i,m} * B^D} \norm{\Phi_{i,m}} } \left( \Phi_{i,m} * B^D * B^D * \sum_{j,n,l} a_{j,n,l} \Phi_{j,n} * \delta_l \right) (k) \\ \nonumber
&= \underset{i,m,k}{\mathrm{argmax}} \frac{1}{\norm{\Phi_{i,m} * B^D} \norm{\Phi_{i,m}} } \left( \Phi_{i,m} * B^D * \delta_k * B^D * \sum_{j,n,l} a_{j,n,l} \Phi_{j,n} * \delta_l \right) (0) \\ 
&= \underset{i,m,k}{\mathrm{argmax}} \sum_{j,n,l} a_{j,n,l} \left\langle \frac{1}{\norm{\Phi_{i,m} * B^D} \norm{\Phi_{i,m}} } \Phi_{i,m} * B^D * \delta_k, B^D * \Phi_{j,n} * \delta_l \right\rangle 
\end{align}
\normalsize
At this point we have to make an approximation. The maximum of the sum is approximately achieved at the maximal summand (this approximation also lies behind the minor loop of standard CLEAN, compare our discussion in Sec. \ref{ssec: clean}). I.e. we solve:
\begin{align} \nonumber
&\underset{i,m,k}{\mathrm{argmax}} \max_{j,n,l} a_{j,n,l} \left\langle \frac{1}{\norm{\Phi_i * B^D} \norm{\Phi_{i,m}} } \Phi_{i,m} * B^D * \delta_k, \Phi_{j,n} * B^D * \delta_l \right\rangle \\ 
&= \underset{i,m,k}{\mathrm{argmax}} \max_{j,n} a_{j,n,k} \left\langle \frac{1}{\norm{\Phi_{i,m} * B^D} \norm{\Phi_{i,m}} } \Phi_{i,m} * B^D,  \Phi_{j,n} * B^D \right\rangle 
\end{align}
where equality holds since $\Phi * B^D$ is centrally peaked.

It is:
\begin{align}
 \langle \Phi_{i,m} * B^D, \Phi_{j,n} * B^D \rangle = 1_{i,j}  \langle \Phi_{i,m} * B^D, \Phi_{i,n} * B^D \rangle,
\end{align}
as the DoB wavelet functions of varying radial widths have distinct supports in the Fourier domain. Hence, we are left with the argmax-problem:
\begin{align}
    &\underset{i,m,k}{\mathrm{argmax}} \max_{n} a_{j,n,k} \frac{1}{\norm{\Phi_{i,m} * B^D} \norm{\Phi_{i,m}} } \langle \Phi_{i,m} * B^D, \Phi_{i,n} * B^D \rangle
\end{align}
Then:
\begin{align} \nonumber
    &\max_{i,m,k,n} a_{j,n,k} \frac{1}{\norm{\Phi_{i,m} * B^D} \norm{\Phi_{i,m}} } \langle \Phi_{i,m} * B^D, \Phi_{i,n} * B^D \rangle \\ \nonumber
    &\leq \max_{i,m,k,n} a_{j,n,k} \frac{1}{\norm{\Phi_{i,m} * B^D} \norm{\Phi_{i,m}} } \norm{\Phi_{i,m} * B^D} \norm{\Phi_{i,n} * B^D} \\ \nonumber
    &= \max_{i,m,k,n\neq N} a_{j,n,k} \frac{\norm{\Phi_{i,n} * B^D} }{\norm{\Phi_{i,m}} } \\
    &= \max_{i,k,n\neq N} a_{j,n,k} \frac{\norm{\Phi_{i,n} * B^D} }{\norm{\Phi_{i,n}} }, \label{eq: relation}
\end{align}
where the last equality holds since $\norm{\Phi_{i,n1}} = \norm{\Phi_{i,n2}}$ for every $n1, n2$. This maximum gets reached exactly for $m = n$. Hence, our selection criterion Eq. \ref{eq: selection} is met by this procedure. 

\section{Reliability of features recovered with DoB-CLEAN} \label{app: reliability}
We now discuss the reliability of the new features recovered in the reanalysis of the BL Lac observations with DoB-CLEAN. Two features are in particular outstanding at highest resolution: the coronal emission around the core, and the possible sign of an edge-brightening. We have demonstrated in Sec. \ref{sec: synthetic_test} that DoB-CLEAN addresses several pathologies of CLEAN, allows for moderate super-resolution and more accurate representation of extended emission. However, as was also demonstrated in Sec. \ref{sec: synthetic_test}, these advantages come to the cost of low-level imaging artifacts, in particular for the more challenging problem of recovering images with super-resolution (i.e. when not convolving with a beam). It is therefore unknown from a-priori how reliable the image features observed with DoB-CLEAN really are. Does DoB-CLEAN resolve some features that were not visible with CLEAN since DoB-CLEAN processes the uv-coverage more seriously? Or does vice versa DoB-CLEAN pick up on some artifacts that were suppressed by CLEAN since the interactive data manipulation (self-calibration, tapering, CLEAN-windows, flagging, ...) is more natural in CLEAN?

\begin{figure*}
    \centering
    \includegraphics[width=\textwidth]{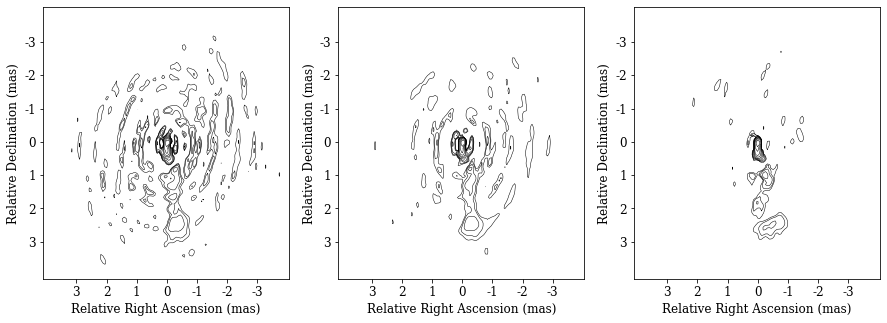}
    \caption{Progress of the cleaning with DoB-CLEAN. Shown is the sum of the recovered image and the residual after 4000 iterations (left panel), 5000 iterations and phase self-calibration (middle panel) and the final ground-only image (right panel). The contour levels are [0.1,0.2,0.4,0.8,1.6,3.2,6.4,12.8,25.6,51.2]\% of the peak brightness.}
    \label{fig: progress}
\end{figure*}

We present in Fig. \ref{fig: progress} the progress of the CLEANing procedure with DoB-CLEAN on the BL Lac data set. The residuals show a ring-like sidelobe structure that indicates the missing of certain scales in the not yet fully converged reconstructions. These scales will be added during later iterations as can be seen from the final reconstruction, i.e. the vanishing residual, in the most-right panel. However, the progress of the DoB-CLEAN procedure highlights a specific requirement: since the image is composed by a sequence of wavelet-subbands that each encode information on a specific spatial scale (i.e. the scale of the ring-like sidelobe pattern, the scale of the second ring-like sidelobe-pattern, ...) a final and clean reconstruction result will be only achievable if the various scales enter the recovered image with the correct weighting relative to each other. If one scale is over-weighted by the reconstruction procedure, the recovered image will contain sidelobes at this spatial scale as well. The correct relative weighting of scales is taken into account by our scale-selection criterion that was proven to be optimal in the absence of calibration issues. However, this fosters an important essential in the application to real, observational data: the self-calibration procedure needs to produce well estimates to the true gains such that no scale will be preferred or suppressed as a consequence of gain variations. Vice versa, the absence of sidelobe emission in the final reconstruction, see the most-right panel in Fig. \ref{fig: progress}, indicates that the calibration and imaging procedure is consistent and was successful.

We note that the coronal emission around the central core component appears at the size of the first sidelobe scale, while the second edge-brightened blob left to the main jet-feature corresponds well to the second sidelobe scale (see Fig. \ref{fig: progress}). This questions the robustness of the presence (DoB-CLEAN) or absence (CLEAN) of these image features. According to our reasoning above, we admit that DoB-CLEAN might be more prone than CLEAN to capture on sidelobe artifacts. However, the overall success of the reconstruction points towards a robust recovery. Moreover, also the CLEAN reconstructions seem to indicate emission to the north-west and south-east of the central core component, comparable to the coronal emission found with DoB-CLEAN, e.g. compare the most right panels in Fig. \ref{fig: clean} and Fig. \ref{fig: full_res} building towards consistence. Typically in CLEAN-like algorithms (such as CLEAN and DoB-CLEAN) the decision which emission is true and which emission is thought to be caused by a sidelobe is answered manually by setting proper CLEAN windows and by self calibration. A final answer, which reconstruction is more correct cannot be given here. We recommend the use of a RML based method that fit the closure quantities instead of the visibilities \citep[e.g.][]{Chael2018, Mueller2022} for consecutive works to build concordance between various reconstructions.

\end{document}